 \documentclass[12pt]{article}
 
 \usepackage{scicite}
 \usepackage{amsmath}
 \usepackage{amsfonts}
 \usepackage{times}
 \usepackage{graphicx}
 \usepackage{textcomp}
 \usepackage{float}
 \usepackage{xcolor}
 \usepackage{soul}
 \usepackage{comment}
 \usepackage{lineno}
 
\newcommand{\be}{\begin{equation}}
\newcommand{\ee}{\end{equation}}
 
 \soulregister\cite{7} 
 \soulregister\ref{7} 
 \soulregister\eqref{7} 
 
 \newenvironment{sciabstract}{%
 	\begin{quote} \bf}
 	{\end{quote}}

 \title{Coherent enhancement of optical remission in diffusive media}

 \author
 {Nicholas Bender,$^{1}$ Arthur Goetschy,$^{2}$ Chia Wei Hsu,$^{3}$\\ 
 Hasan Y\i lmaz,$^{4}$ Pablo Jara Palacios,$^{5}$ \\
 Alexey Yamilov,$^{5\ast}$ and Hui Cao$^{1\ast}$\\
 	\\
 	\normalsize{$^{1}$Department of Applied Physics, Yale University,}\\
 	\normalsize{New Haven, Connecticut 06520, USA}\\
 	\normalsize{$^{2}$ESPCI  Paris,  PSL  University,  CNRS,  Institut  Langevin,}\\
 	\normalsize{1  rue  Jussieu,  F-75005  Paris,  France}\\
 	\normalsize{$^{3}$Ming Hsieh Department of Electrical and Computer Engineering,}\\
 	\normalsize{University of Southern California, Los Angeles, California 90089, USA}\\
 	 \normalsize{$^{4}$Institute of Materials Science and Nanotechnology,}\\
 	\normalsize{National Nanotechnology Research Center (UNAM),}\\
 	\normalsize{Bilkent University, 06800 Ankara, Turkey}\\
 	\normalsize{$^{5}$Physics Department, Missouri University of Science \& Technology,}\\
 	\normalsize{Rolla, Missouri 65409, USA}\\
 	\\
 	\normalsize{$^\ast$ Correspondence should be addressed to hui.cao@yale.edu and/or yamilov@mst.edu.}
 }
 
 \date{}

\begin{document} 
 	
 	
 	\baselineskip24pt
 	
 	
 	\maketitle

 	
 	\begin{sciabstract}
 		
From the earth’s crust to the human brain, remitted waves are used for sensing and imaging in a diverse range of diffusive media. Separating the source and detector increases the penetration depth of remitted light, yet rapidly decreases the signal strength, leading to a poor signal-to-noise ratio. Here, we experimentally and numerically show that wavefront shaping a laser beam incident on a diffusive sample enables an order of magnitude remission enhancement, with a penetration depth of up to 10 transport mean free paths. We develop a theoretical model which predicts the maximal-remission enhancement. Our analysis reveals a significant improvement in the sensitivity of remitted waves, to local changes of absorption deep inside diffusive media. This work illustrates the potential of coherent wavefront control for non-invasive diffuse-wave imaging applications, such as diffuse optical tomography and functional near-infrared spectroscopy.
 		
 	\end{sciabstract}

\section*{Introduction}

The remission geometry is broadly used for imaging and sensing deep inside random media~\cite{yodh1995spectroscopy, campillo2003long, gibson2005recent, durduran2010diffuse, eggebrecht2014mapping, gunther2017review, yucel2017functional, angelo2018review, wapenaar2019green}, because it combines the non-invasiveness of reflection with the deep-penetration of transmission. In real world applications where transmitted waves are either inaccessible or strongly attenuated, diffusive waves remitted from the same side of the medium as their source are collected as signals~\cite{boas2001imaging, ferrari2012brief}. To enhance the penetration depth of detected signals, the source and detector are spatially separated on the medium’s surface. With increasing source-detector distance $d$, waves from the source migrate deeper into the diffusive medium before reaching the detector~\cite{cui1991experimental, feng1995photon}. Their paths are distributed over a banana-shaped region with two ends connecting the source and the detector, with the mid-region reaching deepest (close to $d/2$). Since the waves generated by the source diffuse in all directions, only a small portion eventually reaches the detector. The signal strength decays rapidly with increasing the source-detector separation; thus the signal-to-noise ratio (SNR) is poor for large penetration depths. 

Recent advances in optical wavefront shaping have enabled the control of coherent wave transport in random scattering media, enhancing light transmission and energy deposition~\cite{2012_Mosk_SLM_review, 2017_Rotter_Gigan_review}. Finding the optimal wavefront for an incident beam mostly relies on the detector/camera on the other side or a guide star inside the medium~\cite{horstmeyer2015guidestar, yoon2020deep}. Non-invasive focusing and imaging schemes have been developed utilizing: the optical memory effect~\cite{2012_Bertolotti_Nat, 2014_Katz_NatPhoton, 2015_Yilmaz_Optica, stern2019noninvasive}, nonlinear excitation~\cite{tang2012superpenetration, katz2014noninvasive}, the interaction between light and acoustic waves~\cite{xu2011time, judkewitz2013speckle, chaigne2014controlling, ma2014time, lai2015photoacoustically, katz2019controlling}, and linear fluorescence~\cite{2019_Daniel_OE, 2019_Boniface_Optica, boniface2020non, li2020non}. Moreover, steady-state and time-gated reflection eigenchannels are employed for focusing light onto --and image reconstruction of-- embedded targets~\cite{popoff2011exploiting, choi2013measurement, badon2016smart, jeong2018focusing}. In these setups the backscattered signals, which are collected at the same location as the injected light, have a penetration depth less than or comparable to the transport mean free path $\ell$~\cite{yoon2020deep, Aubry_SciAdv_2020}. Spatial displacement of a source and a detector --the common geometry for diffuse optical tomography (DOT) and functional near-infrared spectroscopy (fNIRS)-- has yet to be explored in wavefront shaping experiments; even though remitted waves can go much deeper than $\ell$ into a diffusive sample~\cite{boas2001imaging, ferrari2012brief}.  

Here we shape the incident wavefront of a monochromatic laser beam to enhance remitted waves at source-detector distances $d \gg \ell$. The key issue is whether the penetration depth is compromised by remission enhancement. We demonstrate, experimentally and numerically, an order of magnitude enhancement of remission with no change in the penetration depth up to 10 $\ell$ deep. Our theoretical model predicts the maximal-remission enhancement and its dependence on the source-detector separation $d$, the transport mean free path $\ell$, and the number of input and output channels. Finally we analyze the sensitivity of remitted waves, to local changes of absorption deep inside diffusive media. The maximum remission eigenchannel enhances the sensitivity by one order of magnitude. This work illustrates the power of wavefront shaping for steering coherent waves deep inside diffusive media, with potential applications in DOT and fNIRS.

\section*{Experimental Setup}

\begin{figure}[H]
     \begin{center}
    	\includegraphics[width=4 in]{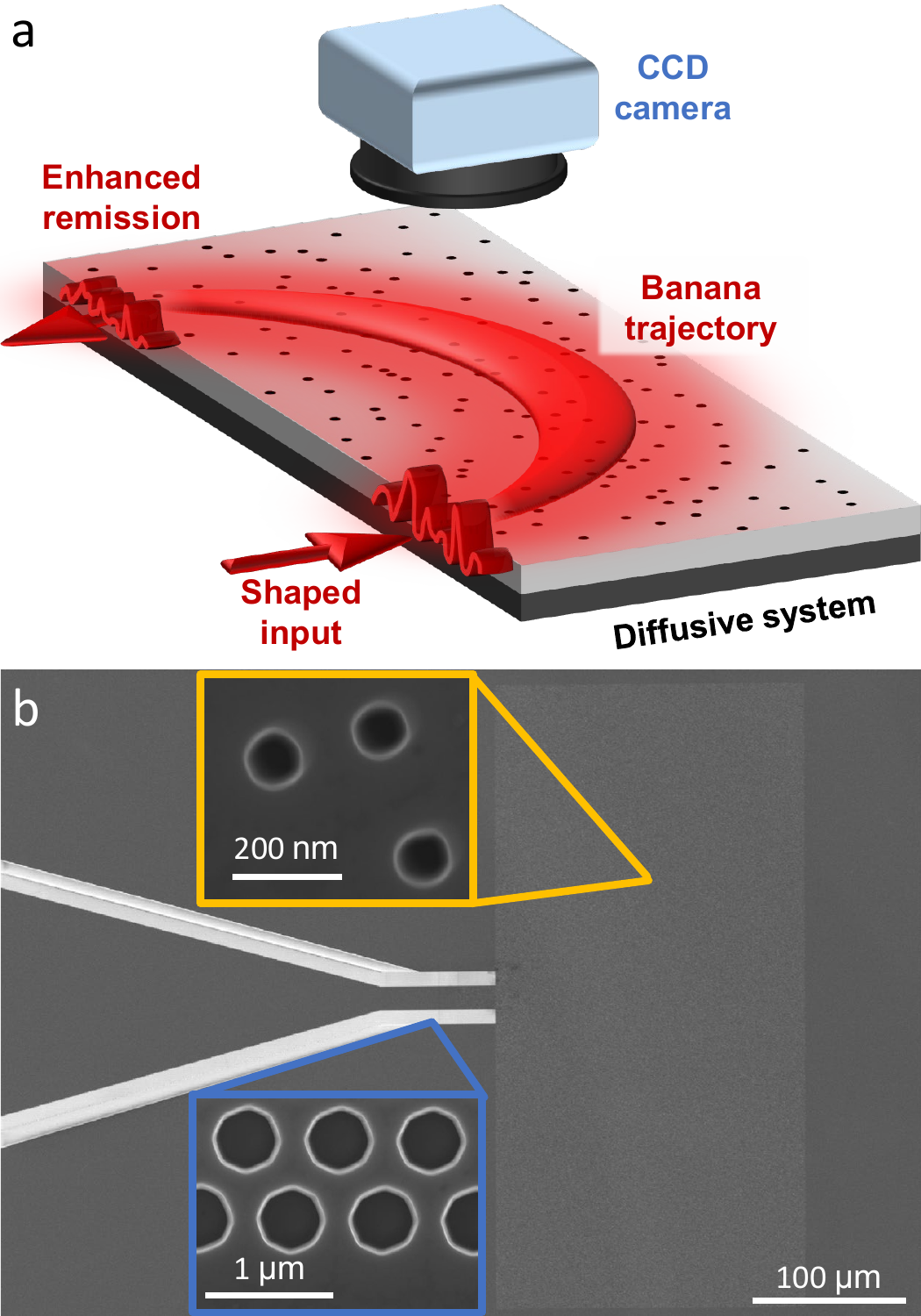}
    	\caption{\label{Figure1} 
    		{\bf Remission enhanced by coherent wavefront control.} (a) A schematic illustrating that shaping the incident wavefront of a laser beam into a 2D diffusive system can steer the light flow along a banana-shaped trajectory to enhance the remission from a preselected site. The internal intensity distribution is recorded by capturing the light scattered out-of-plane with a CCD camera. (b) An example SEM image of the 2D slab consisting of randomly distributed air holes (with a diameter of 100 nm) on a silicon-on-insulator wafer. A monochromatic laser beam is injected through a tapered waveguide (with photonic-crystal sidewalls) into the slab (with open boundaries). 
    	} 
    \end{center}
     \end{figure}

In order to monitor the migration paths of remitted waves inside the diffusive medium, we fabricate two-dimensional (2D) disordered structures on a silicon chip and observe the internal light distribution from the third dimension [Fig.~\ref{Figure1} (a)] \cite{2022_Bender_Deposition}. The remission matrix $\mathcal{R}$ is introduced to relate input fields within a finite region of the interface to remitted waves from another region displaced from the injection site, on the same interface. We experimentally measure $\mathcal{R}$ for different separations from 3$\ell$ to 25$\ell$, find the maximum-remission eigenstates, and investigate their spatial structures.

The 2D diffusive system has a slab geometry (width $W$ = 400 \textmu m, thickness $L$ = 200 \textmu m) and open boundaries on all four sides [Fig.~\ref{Figure1}(b)]. Inside the slab, air holes of 100 nm diameter are randomly distributed with a filling fraction of $2.75\%$. The transport mean free path is $\ell = 6.4$ \textmu m at (vacuum) wavelength $\lambda_0 = 1.55$ \textmu m \cite{2020_Bender_correlations_PRL}. With slab dimensions $L$ and $W$ much larger than $ \ell$ but still smaller than the 2D localization length, light transport is diffusive. Out-of-plane scattering is treated as loss, corresponding to a diffusive dissipation length of $\xi_a$ = 56 \textmu m (see supplementary section 1.1).

A spatial light modulator (SLM) shapes the phase front of the monochromatic laser beam, which is then coupled into a multimode waveguide etched on a silicon-on-insulator wafer. The waveguide delivers light of $\lambda_0 \simeq$ 1.55 \textmu m to a 2D slab on the same chip via 56 guided modes. The field distribution across the entire slab is measured in an interferometric setup. We scan the input wavelength to obtain different configurations.

First, we generate random illumination patterns using the SLM, and map diffusive light transport inside the slab.  Figure~\ref{Figure2}(a) shows the ensemble-averaged intensity distribution $\langle I_{\text{rand}}(y,z)\rangle$. The injection site is centered at $(0, 0)$ and has a width $W_1$ = 15 \textmu m. Away from it, the intensity of the remitted light drops quickly. Along the front boundary $z=0$ of the slab, the diffusive intensity decreases quadratically with distance $|y| = d \gg W_1$. 

The probability of photon migration from the injection site $(0,0)$ to the remission site $(d, 0)$ via the position $(y,z)$ inside the slab equals the product of the probability of migrating from $(0,0)$ to $(y,z)$ and that from $(y,z)$ to $(d, 0)$. The former is proportional to $\langle I_{\text{rand}}(y,z)\rangle$, and the latter to  the average intensity distribution $\langle I^{(d)}_{\text{rand}}(y,z)\rangle$ for light injected at $(d, 0)$, according to optical reciprocity \cite{feng1995photon}. With random incident wavefronts, the maximum probability of photon migration follows a banana-shaped trajectory from $(0, 0)$ to $(d, 0)$. While increasing injection-remission separation gives access to deeper-penetrating light, it comes at the price of a rapidly reduced signal strength.

\section*{Remission Matrix \& Eigenchannels}

Our aim is to utilize the spatial degrees of freedom in the coherent illumination pattern to improve the remitted signal strength. To find the optimal input wavefront, we measure the remission matrix $\mathcal{R}$, and find its associated eigenstates. In a standard DOT setup, light is delivered onto a diffusive sample by a waveguide and the remitted signal is collected by another waveguide. The incident field ${\bf E}_{\rm in}$ and remitted field ${\bf E}_{\rm re}$ are decomposed into $M_{1}$ and $M_{2}$ flux-carrying modes of the waveguides. In a linear scattering medium, they are related by the remission matrix $\mathcal{R}$ as  
\begin{equation}
{\bf E}_{\rm re} = \mathcal{R}  \,{\bf E}_{\rm in}.
\label{remission}
\end{equation} 
Singular value decomposition of $\mathcal{R}$ gives the remission eigenchannels. The one corresponding to the largest singular value has the highest possible remittance. 

While the waveguide mode basis is used in Eq.~\ref{remission}, any orthogonal basis is sufficient. In our experiment, instead of using a waveguide to collect the remitted light, we directly measure the field at the front boundary of the slab at a distance $d$ from the injection waveguide. More specifically, we sample the fields at $20 \times 20$ spatial positions within a 10 \textmu m $\times$ 10 \textmu m square. By displaying orthogonal phase patterns on the SLM, we construct the remission matrix $\mathcal{R}_{\mathrm{SLM} \rightarrow d}$. Singular value decomposition of $\mathcal{R}_{\mathrm{SLM}\rightarrow d}$ provides the remission eigenchannels and associated input vectors. Since our SLM can only modulate phase, not amplitude, it will not excite a pure eigenchannel. Alternatively, we measure the matrix $\mathcal{R}_{\rm SLM\rightarrow SLAB}$ that connects the incoming fields to the field everywhere inside the slab. Multiplying it by  the input vector of a remission eigenchannel recovers the field distribution across the entire slab. 

\begin{figure}[H]
    \begin{center}
    	\includegraphics[width=\linewidth]{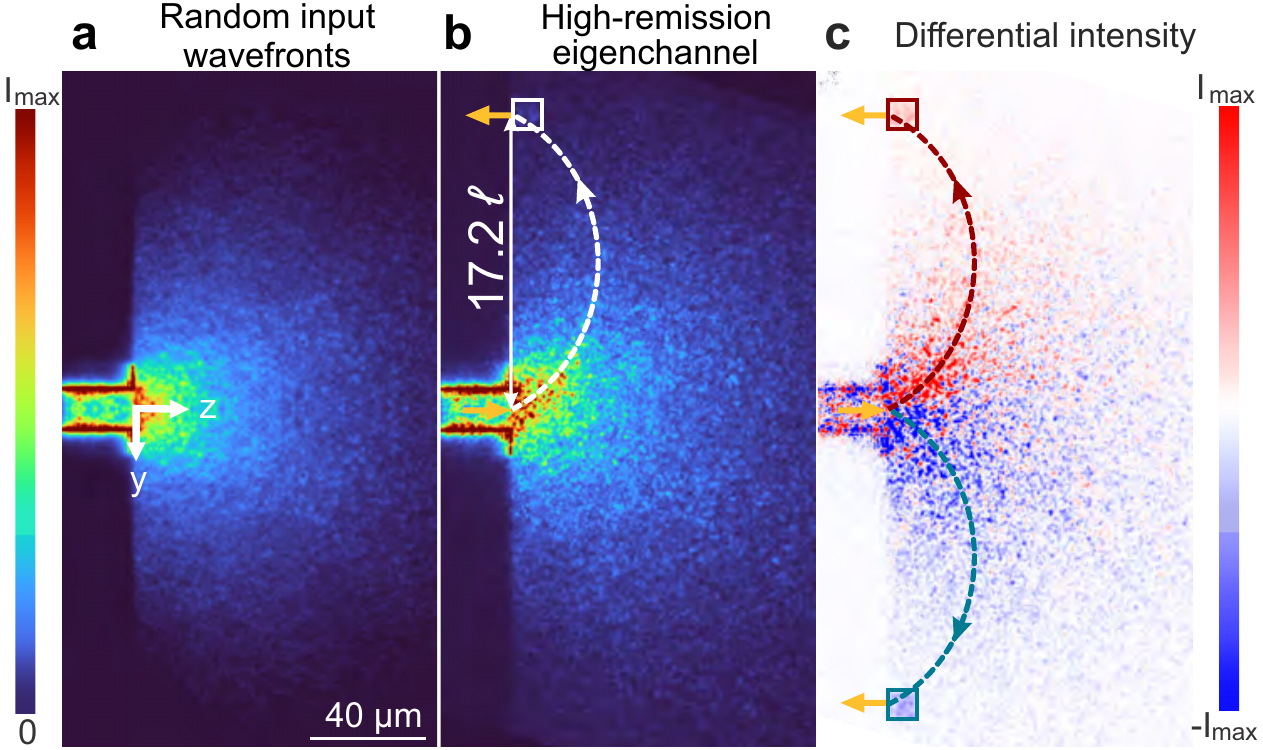}
    	\caption{\label{Figure2} 
    		{\bf High-remission eigenchannel profile.} 
    		(a) Ensemble-averaged intensity distribution, $\langle I_{\text{rand}} (y,z)\rangle$, for random input wavefronts. The injection waveguide is 15 \textmu m wide. 
    		(b) An example high-remission eigenchannel ensemble-averaged intensity distribution $\langle I_{d}(y,z)\rangle$ for the remission site displaced $d = 17.2 \ell$ from the injection site. White dashed curve represents the banana-shaped trajectory for the maximum probability of photon migration with random incident wavefronts. White square denotes 10 \textmu m $\times$ 10 \textmu m area for sampling the resmission.
    		(c) Difference in intensity distribution between high-remission eigenchannels that maximize remission at $y= \pm d$: $\langle I_{+d}(y,z)\rangle-\langle I_{-d}(y,z)\rangle$. It reveals high-remission eigenchannels redistribute energy \textit{inside} the diffusive system along the banana trajectories plotted by dashed lines.  
    		In (a-c), images recorded at 12 different wavelengths, in increments of 1 nm, from 1547 nm to 1558 nm are averaged.
    	} 
    \end{center}
    \end{figure}

Figure~\ref{Figure2}(b) shows an example high-remission eigenchannel profile $\langle I_{d}(y,z)\rangle$. The remission region (white square) is located at $d = 17.2 \ell$ from the input waveguide. $\langle I_{d}(y,z)\rangle$ shows that the diffuse light is steered towards the detector via the banana trajectory, which is obtained by tracing the maximum photon migration probability under random illumination (see supplementary section 1.3). The steering is further illustrated by the difference $\langle I_{+d}(y,z)\rangle-\langle I_{-d}(y,z)\rangle$ in Fig.~\ref{Figure2}(c). The positive (red) or negative (blue) intensity pattern reveals light is directed towards the remission site at $(d, 0)$ or $(-d, 0)$. In both cases, high-remission eigenchannels redistribute optical energy along the banana trajectories connecting the injection and detection sites. 

\begin{figure}[ht]
    \begin{center}
    	\includegraphics[width=\linewidth]{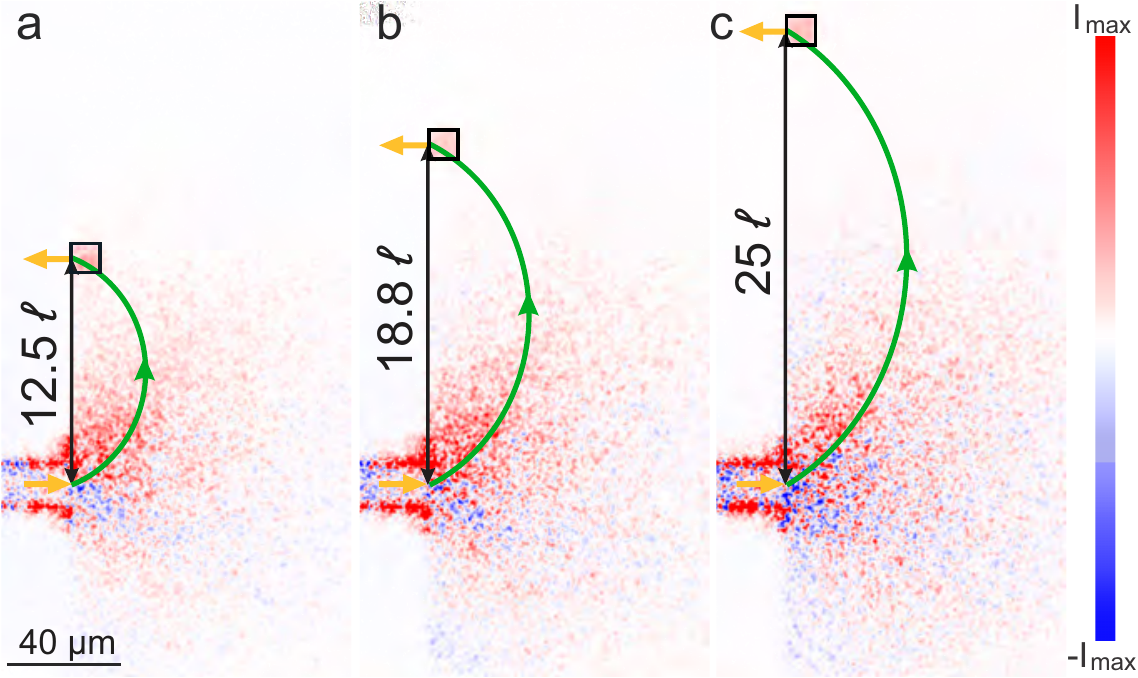}
    	\caption{\label{Figure3} 
    		{\bf Difference between high-remission eigenchannel profiles and random illumination patterns.} 
    		Ensemble-averaged intensity distributions for high-remission eigenchannels subtracted by those of random input wavefronts $\langle I_{d}(y,z)\rangle-\langle I_{\text{rand}}(y,z)\rangle$ for remission sites (black square) located at (a) $d = 12.5 \ell$, (b) $d = 18.8 \ell$, (c) $ d = 25 \ell$ away from the injection waveguide center. The solid-green lines show the banana trajectory of photon migration for each separation. In (a-c) data taken over 12 different wavelengths, in increments of 1 nm, from 1547 nm to 1558 nm are averaged.
    	} 
    \end{center}
    \end{figure}

We vary the injection-detection distance $d$ and plot the difference between high-remission eigenchannel profile $\langle I_{d}(y,z)\rangle$ and random input pattern $\langle I_{\text{rand}} (y,z)\rangle$ for $d = 12.5 \ell,$ $18.8 \ell$, $25.0 \ell$ in Fig.~\ref{Figure3}. The positive (red) and negative (blue) intensity areas demonstrate that high-remission eigenchannels redistribute optical energy inside the system compared to random inputs. Furthermore, the positive (red) intensity regions in (a-c) concentrate along the banana trajectories (green line). Therefore, it is possible to direct the flow of light deep inside diffusive systems by coupling into high-remission eigenchannels. The penetration depth is not affected at all by the remission enhancement. 

In our experiment, wavefront shaping can also increase the out-of-plane scattering from the detection site, which inflates the remission intensity. We thus resort to numerical simulations of the experimental system to quantify the remission enhancement as a function of $d$. For this purpose, we simulate wave propagation in 2D disordered slabs using the Kwant package \cite{2014_Groth_Kwant,2016_Sarma_Open_Channels}, and compute the remission matrix $\mathcal{R}$ with $M_{1}=56$ input channels and $M_{2}=37$ output channels. The eigenvalues $\rho$ of $\mathcal{R}^\dagger\mathcal{R}$ give the fraction of power remitted to the output waveguide of width $W_2 = 10$ \textmu m when sending the associated input vectors into the slab. We represent in Fig.~4(a) the probability density function (PDF) of non-zero eigenvalues, $P(\rho)$, for a broad range of source-detector distance $d$  (larger than the transport mean free path $\ell=6.4$ \textmu m). The PDF decays monotonically, indicating that most eigenstates deliver little power at the remission port. However, we note that the PDF presents an upper edge $\rho_{\text{max}}$ much larger than $\rho_{\text{rand}}$ (the fraction of power delivered by random input illumination). For example, $10\%$ of the total injected power can be remitted at a distance $d\simeq 8\, \ell$, compared to $1\%$ for random illumination.
As the distance $d$ increases, all eigenvalues get smaller, since less power is collected, and the PDF narrows. To quantify the benefit of using the eigenstate associated with the largest remission eigenvalue instead of random illumination, we represent in Fig.~4(b) the ratio $\rho_{\text{max}}/\rho_{\text{rand}}$ as a function of the distance $d$. Enhancement typically larger than $10$ is reached (blue circles). Remarkably, the enhancement $\rho_{\text{max}}/\rho_{\text{rand}}$ increases with $d$, which illustrates the power of coherent wavefront control for $d\gg \ell$. When including out-of-plane scattering loss in our simulations (purples squares), the enhancement $\rho_{\text{max}}/\rho_{\text{rand}}$ slightly increases because dissipation has more impact on the random input propagation than on a high-remission eigenchannel \cite{liew2014transmission}; see supplementary section 2.4 for the full distributions $P(\rho)$ in the presence of loss. Moreover,  increasing the scattering strength of the disordered medium through a reduction of $\ell$ (red triangles) leads to further enhancement of remission $\rho_{\text{max}}/\rho_{\text{rand}}$, which can be as large as $20$ at $d \simeq 47 \ell = 150$ \textmu m.
 
To elucidate the intricate dependence of $P(\rho)$ and $\rho_{\text{max}}/\rho_{\text{rand}}$ on relevant parameters $d$, $\ell$, $M_{1}$, and $M_{2}$, we develop a theoretical model based on a combination of random matrix theory and microscopic computations of intensity fluctuations in remission. Our approach relies on the concept that any structure consisting of effective diffusive systems with comparable conductance in series is characterized by a universal bimodal eigenvalue distribution, irrespective of the microscopic origin of scattering and the geometry of the scattering system~\cite{2009_Nazarov_Book_Quantum_Transport}. This distribution, initially put forward for transmission through diffusive wires~\cite{1997_Beenakker}, can in principle be used for the remission configuration, as long as input and output spatial channels do not belong to the same waveguide. However in our setup, light propagates in an open geometry with input/output covering only a small fraction of the total surface area. Therefore, we must also take into account the incomplete channel control of the injection and detection. We model the remission matrix $\mathcal{R}$ as a filtered matrix of dimension $M_{2} \times M_{1}$, drawn from a virtual $M_0\times M_0$ matrix characterized by a bimodal distribution of eigenvalues with mean $\bar{\rho}_0$, and use the predictions of the filtered random matrix (FRM) ensemble~\cite{2013_Goetschy}. The only free parameters of this model are thus $M_0$ and $\bar{\rho_0}$, which can be determined from microscopic calculations of the first two moments of the distribution $P(\rho)$. Details of the full model are given in supplementary section 2. Solid lines in Fig.~4(a,b) show our theoretical predictions for $P(\rho)$ and its upper edge $\rho_{\text{max}}$, which are in excellent agreement with the numerical results.

Next we consider limiting cases. If the number of output spatial channels $M_2$ is equal to 1, the remission enhancement $\rho_{\text{max}}/\rho_{\text{rand}}$ equals the number of input channels $M_1$, regardless of the injection-remission distance $d$ and the transport mean free path $\ell$. As $M_2$ increases,
the maximal remitted signal $\rho_{\text{max}}$ grows, but 
the enhancement $\rho_{\text{max}}/\rho_{\text{rand}}$ drops.  
A key quantity controlling the scaling of  $\rho_{\text{max}}/\rho_{\text{rand}}$ with microscopic parameters is the non-Gaussian component of intensity fluctuations measured 
at the remission port and generated by random illumination from the injection port. These fluctuations are commonly termed $C_2$~\cite{1988_Feng_PRL}; see supplementary section 2.2 for their explicit calculation. When  $C_2$ is small, $M_2 < 1/C_2$, the remission matrix $\mathcal{R}$ can be approximated by a Gaussian random matrix, and the enhancement factor $\rho_{\text{max}}/\rho_{\text{rand}}$ scales as  $\sim M_1/ M_2$~\cite{2013_Goetschy,2014_Popoff, hsu2017correlation}. However, if $C_2$ is larger, $M_2 > 1/ C_2 $, non-Gaussian intensity correlations can further enhance the remission. In a 2D diffusive system, $C_2$ leads to the increase of $\rho_{\text{max}}/\rho_{\text{rand}}$  with both scattering strength $1/k\ell$ and injection-remission distance $d$.
Indeed, in the situation $d\gg \ell$ we find that the remission enhancement depends on a single parameter: the normalized variance $\text{Var}(\rho/\left<\rho \right >)$ of the PDF $P(\rho)$,
 related to $C_2$ as $\text{Var}(\rho/\left<\rho \right >)=M_{2} C_2 +M_{2}/M_{1}$, where $C_2 \approx \text{ln}(d/W_1)/k\ell$ 
 (see supplementary section 2.2). In the limit $M_{2} \gg 1/C_2 $, the remission enhancement takes the form (supplementary section 2.3) 
\begin{equation}
\frac{\rho_{\text{max}}}{\rho_{\text{rand}}} \simeq \frac{3}{2}M_{1}C_2 \propto M_{1} \frac{\text{ln}(d/W_1)}{k\ell}.
\end{equation}
Contrary to the remission under random illumination $\rho_{\text{rand}} \simeq M_{2} k\ell /(kd)^2$, the high-remission eigenchannel generates a flux $\rho_{\text{max}} \simeq M_1 M_{2} \, \text{ln}(d/W_1) /(kd)^2$ that is independent of the scattering strength $k\ell$. Ignoring the weak dependence of $\text{ln}(d/W_1)$ on $M_1$, the enhancement factor scales linearly with the number of input channels $M_{1}$. Furthermore, the dependence of $C_2$ on $d$ and $k\ell$ explains the general trends beyond the above limits in Fig.~4(b). We refer to the  supplementary section 4.3 for a study of the continuous evolution of $\rho_{\text{max}}/\rho_{\text{rand}}$ with $M_2$. 

\section*{Sensitivity Analysis}

Given that high-remission eigenchannels improve the signal-to-noise ratio, a natural question is whether they provide higher sensitivity to local perturbations of the dielectric constant inside a diffusive medium. The answer is important to DOT and fNIRS, which often monitor the change in remitted signal due to localized absorptive targets. The answer to this question is not straightforward because sensitivity depends not only on the value of remission, but also on position inside the medium, as shown below. Furthermore, prior analysis of the problem, based on the diffusion equation~\cite{feng1995photon}, is not applicable as the enhanced remission here is achieved through wave interference: which is not captured by diffusion theory.

Let $R$ be the total power collected at the remission port divided by the incident power for an arbitrary incident field $E_{\rm in}$. With $E_{\rm in}$ fixed, a small amount of absorption is introduced as the imaginary part of the relative permittivity $d\varepsilon_i$ over a subwavelength area $A_{\varepsilon}$ centered at location ${\bf r}_0$. It changes the collected remission by $dR$.
The sensitivity is defined as $S \equiv - dR/d\varepsilon_i$.
Under the scalar wave equation approximation in 2D, we show in supplementary section 3 that
\begin{equation}
\label{eq:sensitivity}
S({\bf r}_0; E_{\rm in})
\equiv - \left. \frac{dR\left[E_{\rm in}\right]}{d\varepsilon_i} \right\vert_{{\bf r}_0}
= k_0^2 A_{\varepsilon} \frac{ {\rm Re} \left[ E_t({\bf r}_0) E_c({\bf r}_0) \right]}{ \int dy \, {\rm Im} \left[ E_{\rm in}^*{\partial E_{\rm in}}/{\partial z} \right]_{z=0}},
\end{equation}
where $k_0 = \omega/c$ is the vacuum wave number, $E_t$ is the total field given $E_{\rm in}$ as the incident field, and $E_c$ is the total field with $E_t^*$ in the remission port as the incident field.
Eq.~\eqref{eq:sensitivity} generalizes the adjoint method commonly used in inverse designs~\cite{2018_Molesky_nphoton_review} to multi-channel systems. We evaluate the sensitivity $S({\bf r}_0; E_{\rm in})$ for different input wavefront $E_{\rm in}$ in our numerical simulation.

Figure \ref{Fig_sensitivity}(a-b) shows the ensemble-averaged sensitivity map $S({\bf r}_0; E_{\rm in})$ computed using Eq.~\eqref{eq:sensitivity} for random input wavefronts and for high-remission eigenchannels respectively, in a lossless system.
The sensitivity map of high-remission eigenchannels has the same spatial profile as that of random inputs, with the sensitivity maximized along the banana trajectory.
The high-remission eigenchannels improve the sensitivity about 11 times along such trajectory in Fig.~\ref{Fig_sensitivity}(c).
We further find that the sensitivity enhancement increases with $d$, similar to the remission enhancement. Notably, the sensitivity enhancement is even larger than the remission enhancement in Fig.~\ref{Fig_sensitivity}(d).
To illustrate that the sensitivity depends not just on the remission, we separate the incident and the scattered contributions of the conjugate field, $E_c = E_c^{\rm in} + E_c^{\rm sca}$.  The numerator of Eq.~\eqref{eq:sensitivity} has two terms: ${\rm Re} \left[ E_t E_c^{\rm in} \right]$ and ${\rm Re} \left[ E_t E_c^{\rm sca} \right]$.
A high-remission eigenchannel naturally enhances the first term which is proportional to the remission $R$ for ${\bf r}_0$ near the remission port, but it may also increase the second term to further enhance the sensitivity.
Similar results are observed in systems with loss (Supplementary Fig.~\ref{Sup:Figure4}).

\section*{Discussion \& Conclusion}


We have shown that coherent wavefront shaping greatly enhances the remitted signal and its sensitivity to local change of absorption deep inside a diffusive medium. Our method differs from the existing method of structured illumination in optical tomography, which utilizes an incoherent light and modulates only its intensity \cite{ angelo2018review}. While the latter improves the speed and accuracy of image reconstruction, it does not increase the remitted signal \cite{lukic2009optical, konecky2009quantitative}. In our case, coherent light must be used for illumination, both its amplitude and phase are modulated. The phase modulation is essential to the enhancement of remitted signal via constructive interference of multiply scattered light. 

While the current study is conducted on 2D diffusive systems, our method is applicable to 3D, so is our theoretical model. The remission enhancement increases with $M_1$, but decreases with $M_2$. A difference from 2D is that the maximal-remission enhancement in 3D no longer varies with $d$, because $C_2$ becomes independent of $d$ (supplementary section 2.3). However, the $1/\ell$ dependence due to $C_2$ is preserved. In DOT and fNIRS, the commonly used 3D biological samples have negligible $C_2 \sim 1/ {\sqrt{M_1}k \ell} \ll 1/M_2$, and the maximal-remission enhancement is approximately $\rho_{\text{max}}/\rho_{\text{rand}} \simeq \left( 1 + \sqrt{M_1/M_2} \right)^2$, according to the Marcenko--Pastur law~\cite{hsu2017correlation}.      

In conclusion, we have introduced and investigated, both theoretically and experimentally, the concept of remission eigenchannels in open  diffusive systems with a slab geometry. Using our on-chip interferometric platform, we measure remission matrices and investigate the associated eigenchannel profiles inside disordered systems. Selective excitation of high-remission eigenchannels significantly enhances the signal strength of diffusive remitted light with deep penetration up to 10 transport mean free paths. The greatly improved sensitivity of remitted signals, to local perturbations deep inside diffusive media, is promising for many diffusive-wave imaging and sensing applications, ranging from seismology to noninvasive photomedical devices and brain-computer interfaces.

 	\bibliography{Citations}
 	
 	\bibliographystyle{Science}
 	
\section*{Acknowledgments}
H.C. and A.Y. thank Michael A. Choma for stimulating discussions. 
{\bf Funding}: This work is supported partly by the Office of Naval Research (ONR) under Grant No. N00014-20-1-2197, by the National Science Foundation under Grant Nos. DMR-1905465, DMR-1905442, OAC-1919789, and ECCS-2146021, and by LABEX WIFI (Laboratory of Excellence within the French Program “Investments for the Future”) under references ANR-10-LABX-24 and ANR-10-IDEX-0001-02 PSL*.

{\bf Author contributions}: N.B. conducted the experiments and analyzed the data; A.Y. performed the numerical simulations and derived the long-range correlations with P.J.P. A.G. developed the analytical model. C.W.H. did the sensitivity analysis with A.G. and A.Y. H.Y. contributed to the experimental study. H.C. initiated the project and supervised the research. All authors contributed to the manuscript writing and editing. 

{\bf Competing interests}: The authors
declare no competing financial interests. 

{\bf Data and materials availability}: All data needed to evaluate the conclusions in the paper are present in the paper or the supplementary materials.
 	
     \newpage
     
     \newpage
\newpage
\newpage

    \begin{figure}[ht]
    \begin{center}
    	\includegraphics[width=\linewidth]{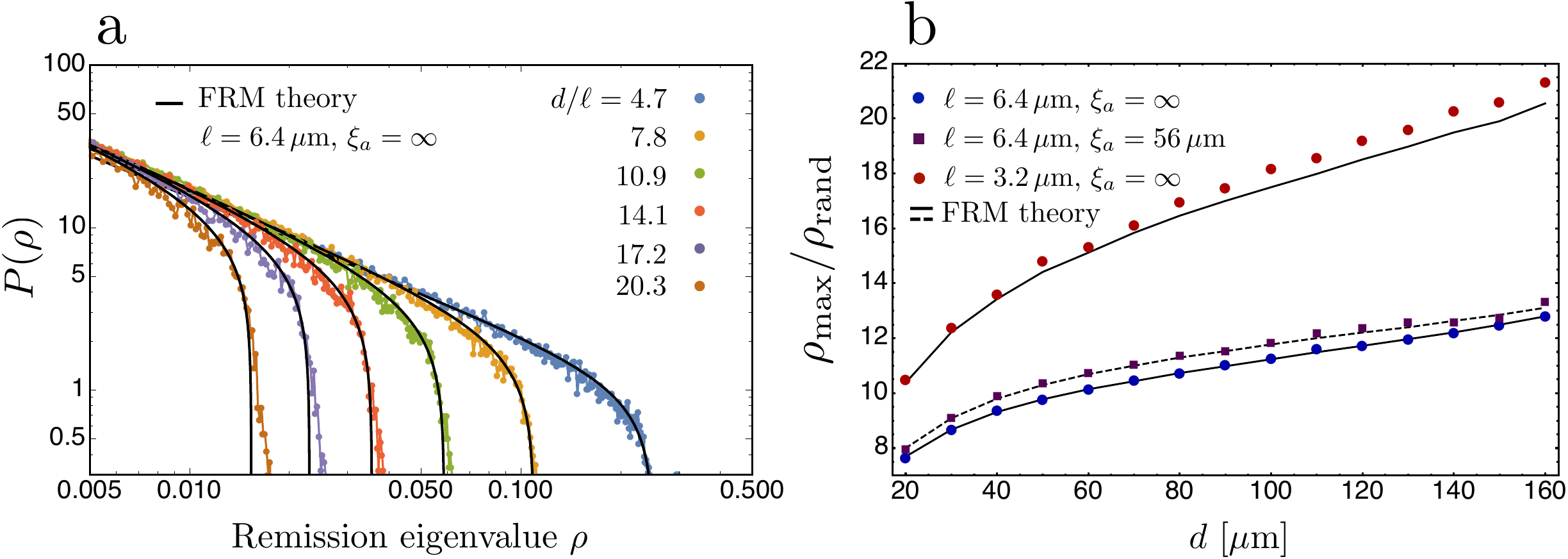}
    	\caption{\label{Fig_theory} 
    	{\bf Theoretical model and numerical simulation of maximal remission enhancement.}
    	(a) Probability density $P(\rho)$ of remission eigenvalue $\rho$ for varying source-detector distance $d$ normalized by the transport mean free path $\ell$ = 6.4 \textmu m in a 2D lossless diffusive slab. Analytical predictions (solid lines) agree with simulation results (dots) averaged over $2275$ disorder configurations.
    	(b) Maximal remission enhancement $\rho_{\text{max}}/\rho_{\text{rand}}$ for three diffusive systems with different amounts of scattering and loss (see legend). $\rho_{\text{max}}/\rho_{\text{rand}}$ increases with $d$. Shorter $\ell$ leads to stronger remission enhancement. With loss ($\xi_a$ = 56 \textmu m) $\rho_{\text{max}}/\rho_{\text{rand}}$ is slightly larger than that without loss ($\xi_a$ = $\infty$). 
    	The input waveguide has a width $W_1$ = 15 \textmu m and support $M_1=56$ modes, and the output waveguide has  $W_2$ = 10 \textmu m and $M_2=37$. } 
    \end{center}
    \end{figure}

\newpage
    \begin{figure}[H]
    \begin{center}
    	\includegraphics[width=4in]{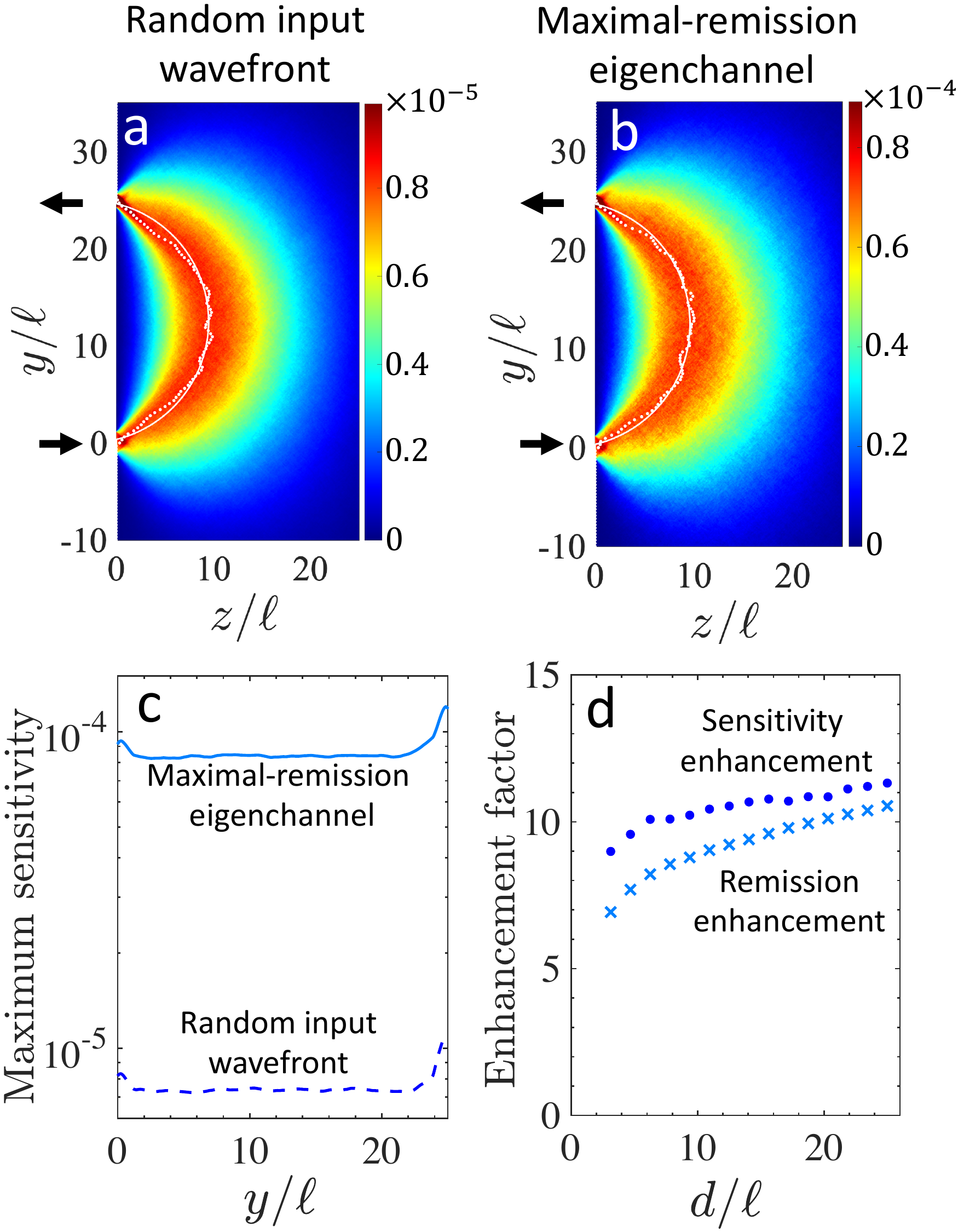}
    	\caption{\label{Fig_sensitivity} 
    	{\bf Sensitivity enhancement by maximal-remission eigenchannel.} 
    	Numerically calculated sensitivity of remission, i.e., change in the remitted signal due to local absorption inside a 2D lossless diffusive system, for random input wavefronts (a) and maximal-remission eigenchannel (b). The parameters are $W_1$ = 15 \textmu m,  $W_2$ = 10 \textmu m, $M_1$ = 56, $M_2$ = 37, $k_0 A_{\varepsilon}^{1/2} =0.65$, $\ell$ = 6.4 \textmu m, and $\xi_a$ = $\infty$. Average over 1000 disorder realizations is performed. White dots denote the depth where the maximum sensitivity is reached for a given value of $y$, fitted by part of a circle -- dashed line -- to show the banana trajectory. The sensitivity map is identical in (a) and (b), confirming the penetration depth is not compromised by the enhanced remission.  
    	(c) Maximum sensitivity vs. $y$ from (a,b) showing  an order of magnitude enhancement by the maximal-remission eigenchannel.
    	(d) Sensitivity enhancement at $y=d/2$ (circles) compared to remission enhancement (crosses) as a function of injection-remission separation $d$.
    	} 
    \end{center}
    \end{figure}
    \newpage

 	\section*{Supplementary materials}

 	\section{Experiment}
 	
 	\subsection{Diffusive Samples}
 	
\begin{figure}
    \begin{center}
    	\includegraphics[width=4 in]{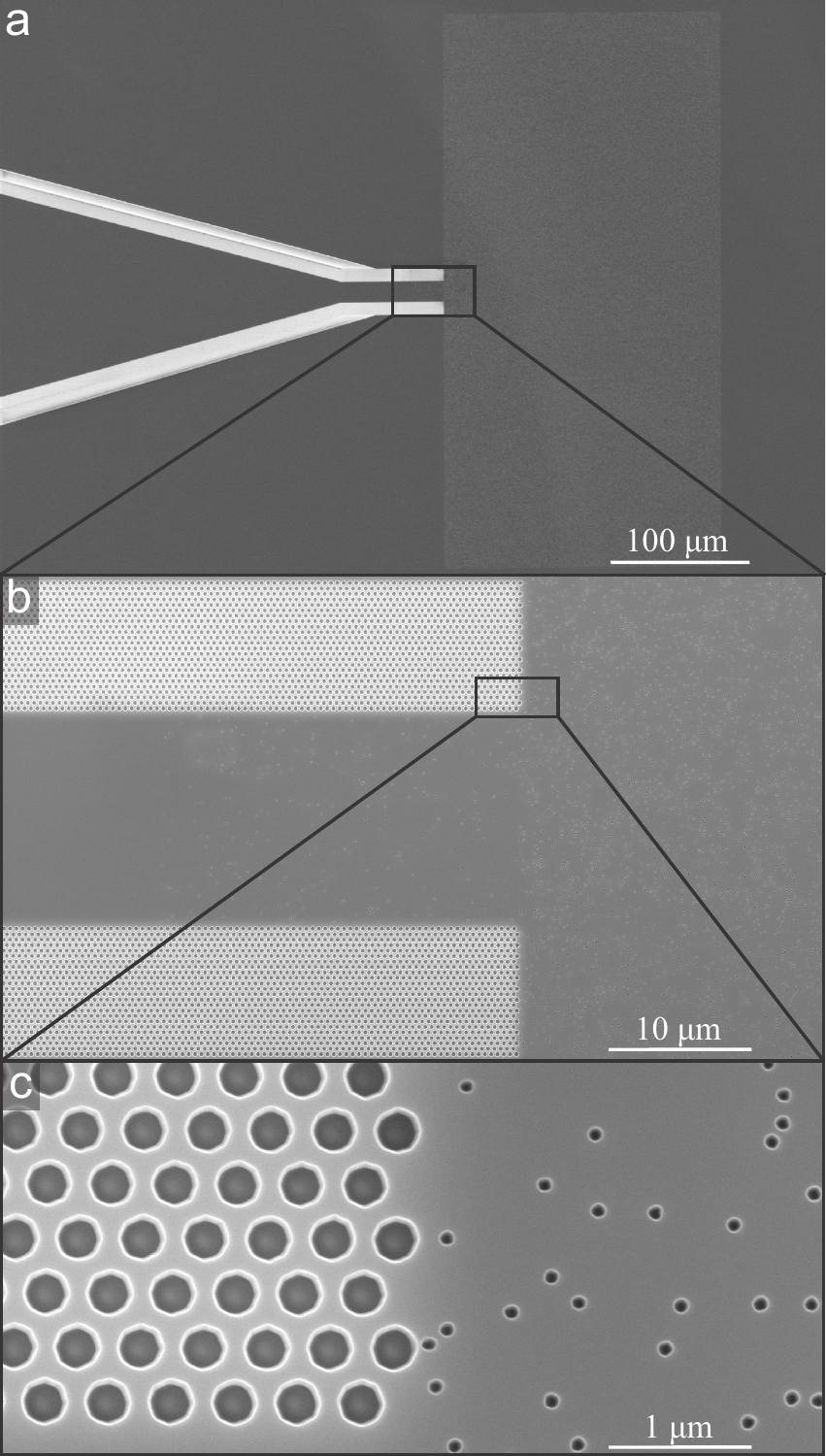}
    	\caption{\label{Sup:Figure1} 
    		{\bf 2D diffusive sample of slab geometry.} 
    		(a) SEM image of the disordered slab and a tapered waveguide for light injection, fabricated on a silicon-on-insulator wafer. 
    		(b) Magnified SEM image showing the 15-\textmu m-wide input waveguide at the disordered slab's front interface.
    		(c) Further magnification revealing a trigonal lattice of air holes (radius = 155 nm, lattice constant = 440 nm) in the waveguide sidewall and examples of the randomly distributed 100-nm-diameter holes inside the slab.  
    	} 
    \end{center}
    \end{figure}

    To directly observe the spatial structure of remission eigenchannels \textit{inside} open diffusive systems, we fabricate two-dimensional diffusive structures on a silicon-on-insulator (SOI) wafer using electron beam lithography and reactive ion etching. An example of the fabricated structure is shown in Fig.~\ref{Sup:Figure1}. The on-chip optical structures we use to route light into the diffusive slab are described in the following paragraphs. 

    The probe light is injected from the side/edge of the wafer into a ridge waveguide (width = 300 \textmu m, length = 15 mm). It then enters a tapered waveguide (tapering angle = $15^{\circ}$), shown in Fig.~\ref{Sup:Figure1}(a). The tapered waveguide width decreases gradually from 300 \textmu m to 15 \textmu m. The tapering results in waveguide mode mixing. To avoid significant light leakage, the sidewalls of the tapered waveguide consist of a trigonal lattice of air holes (radius = 155 nm, lattice constant = 440 nm). They provide a 2D complete bandgap for TE polarized light (electric field parallel to the sample surface) within the wavelength range of 1120 nm to 1580 nm. In addition, we etch a 10 \textmu m trench on the outside interface of the photonic crystal layer. The trench scatters light out of the wafer, preventing any light leaking through the photonic-crystal layer from reaching the open diffusive slab. Because the remitted light can be orders of magnitude weaker than the incident light, mitigating any stray light from the input is essential.
    
    As can be seen in Fig.~\ref{Sup:Figure1}(b), we include a ``buffer’’ region in the injection waveguide, after the taper segment, consisting of a small number of random holes~\cite{2020_Bender_correlations_PRL}. The buffer region causes mode mixing in the lead waveguide, ensuring the incident light is distributed over all available spatial channels. 

    We inject light into a diffusive system configured in a slab geometry with a width of $W$ = 400 \textmu m, a depth of $L$ = 200 \textmu m, and open boundaries on all sides. Inside the diffusive slab, air holes (diameter = 100 nm) induce light scattering and are randomly distributed with a minimum (edge-to-edge) distance of 50 nm, shown in Fig.~\ref{Sup:Figure1}(c). The filling fraction of air holes is $2.75\%$ and as a result, the transport mean free path $\ell = 6.4$ \textmu m and diffusive dissipation length $\xi_a$ = 56 \textmu m~\cite{2020_Bender_correlations_PRL}. In-plane light scattering by a subwavelength air hole is isotropic, and therefore the transport mean free path is equal to the scattering mean free path.

    \subsection{Experimental Setup}
    
    A detailed schematic of our experimental setup is presented in Fig.~\ref{Sup:Figure2}. A wavelength tunable laser (Keysight 81960A) outputs a linearly-polarized continuous-wave (CW) beam with a wavelength around 1554 nm. The collimated beam is split with a polarizing beam splitter into two beams. One is used as a reference beam --with its polarization subsequently flipped with a half-wave plate-- while the other illuminates a phase-only SLM (Hamamatsu LCoS X10468). A one-dimensional (1D) phase-modulation pattern consisting of 128 macropixels is displayed on the SLM. Each macropixel has $4\times800$ SLM pixels. We image the field reflected from the SLM plane onto the back focal plane of a long-working-distance objective Obj. 1 (Mitutoyo M Plan APO NIR HR100$\times$, Numerical Aperture = 0.7) using two lenses with focal lengths of $f_{1}=400$ mm and $f_{2}=75$ mm. To prevent the unmodulated light from entering the objective lens, we display a binary diffraction grating within each macropixel to shift the modulated light away from the unmodulated light in the focal plane of the first lens $f_{1}$. With a slit in the focal plane, we block all light except the phase-modulated light in the first diffraction order. Right after the slit and before the second lens $f_{2}$, we insert a half-wave plate to rotate the polarization of light so that it is transverse-electric (TE) polarized relative to our sample (electric field parallel to the sample surface). The waveguide entrance at the edge of our SOI wafer is placed at the front focal plane of Obj. 1, so that it is illuminated with the Fourier transform of the phase-modulation pattern displayed on the SLM. From the top of the wafer, another long-working-distance objective Obj. 2 (Mitutoyo M Plan APO NIR HR100$\times$) collects light scattered out-of-plane from the sample. We use a third lens with a focal length of $f_{3}=100$ mm together with Obj. 2 to image the light distribution inside the sample. In conjunction, the lens and the objective magnify the sample image by $50$ times. Using a second beam splitter, we combine the signal from the sample with the reference beam. Afterwards, a polarizing beam splitter ensures that the signal has the same polarization as the reference beam. The resulting interference patterns produced by the reference and signal are recorded with an IR CCD camera (Allied Vision Goldeye G-032 Cool).   
    
 	\subsection{Remitted Light For Random Illumination}
     	
    When a finite region of the disordered slab’s front interface is illuminated with a random wavefront, the light diffuses inside the slab in every direction. The experimentally measured, ensemble-averaged intensity distribution, $\langle I_{\text{rand}}(y,z)\rangle$, generated by random illumination patterns is shown in Fig.~\ref{Sup:Figure3}(a).  
    The intensity distribution $\langle I^{(d)}_{\text{rand}}(y,z)\rangle$ for light injected at $(d, 0)$ is obtained by shifting $\langle I_{\text{rand}}(y,z)\rangle$ along $y$-axis by a distance $d$. The product $\langle I_{\text{rand}}(y,z)\rangle \times \langle I^{(d)}_{\text{rand}}(y,z)\rangle$ is proportional to the probability of photon migrating from the injection site $(0,0)$ to the remission site $(d, 0)$ via the position $(y,z)$ inside the diffusive system, as shown in Fig.~\ref{Sup:Figure3}(b). For every cross-section of $y$ between $0$ and $d$, we locate the maximum probability, max$[\langle I_{\text{rand}}(y,z)\rangle \times \langle I^{(d)}_{\text{rand}}(y,z)\rangle]_{z}$, and fit the corresponding location to part of a circle (purple line) to obtain the most-probable trajectory of the remitted light, for a given injection-remission separation $d$. In Fig.~\ref{Sup:Figure3}(c) we plot such trajectory as a function of $d$. The penetration depth of the remitted light increases with $d$. In order for photons to reach the remission site further away from the injection site, they must travel deeper inside the diffusive system to avoid leaking through the front interface ($z=0$). While increased injection-remission distance gives access to a larger penetration depth, it comes with the price of a $1/d^2$ reducion of signal-strength: $\langle I_{\text{rand}}(y=\pm d, z=0)\rangle \propto 1/d^2$. Our goal is to use coherent wavefront shaping to manipulate the spatial degrees of freedom, in the incident wavefront, to improve the remitted signal strength without compromising the penetration depth.
        
    \subsection{Remission Matrix Measurement}
\begin{figure}
    \begin{center}
    	\includegraphics[width=\linewidth]{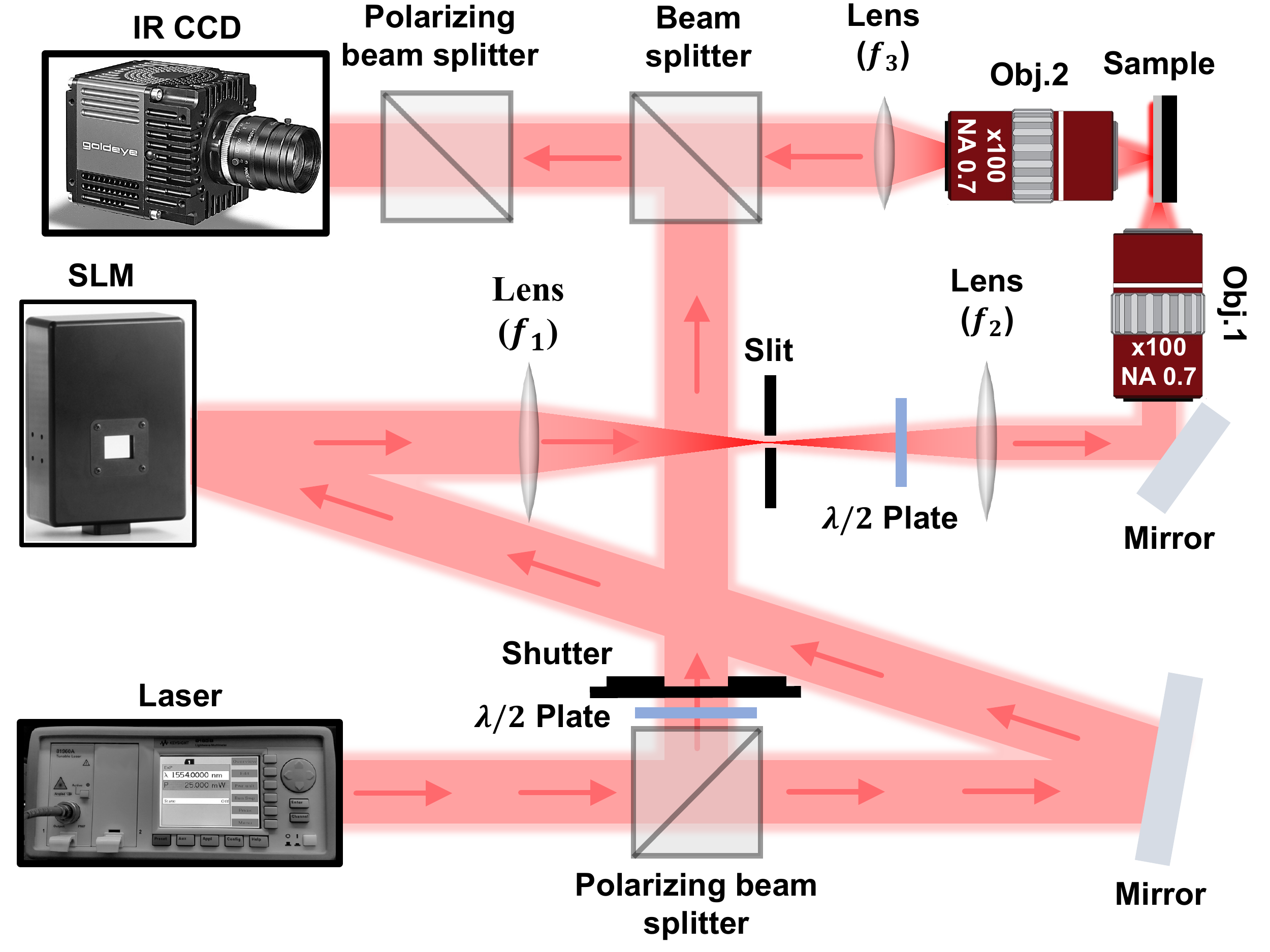}
    	\caption{\label{Sup:Figure2} 
    		{\bf A schematic of our optical setup.} 
 				Monochromatic light from a tunable CW laser is split into two linearly polarized beams. One beam illuminates the phase modulating region of a spatial light modulator (SLM). The other beam is used as a reference. The SLM controls the input wavefront injected to the on-chip waveguide. A beam splitter merges the light collected from the top of our sample with the reference beam, and their interference pattern is recorded by an IR CCD. The focal length of the three lenses in this setup are: $f_{1}=400$ mm, $f_{2}=75$ mm, and $f_{3}=100$ mm. Obj. 1 and Obj. 2 are 100$\times$ objective lens of NA = 0.7. 
    	} 
    \end{center}
    \end{figure}

\begin{figure}
    \begin{center}
    	\includegraphics[width=\linewidth]{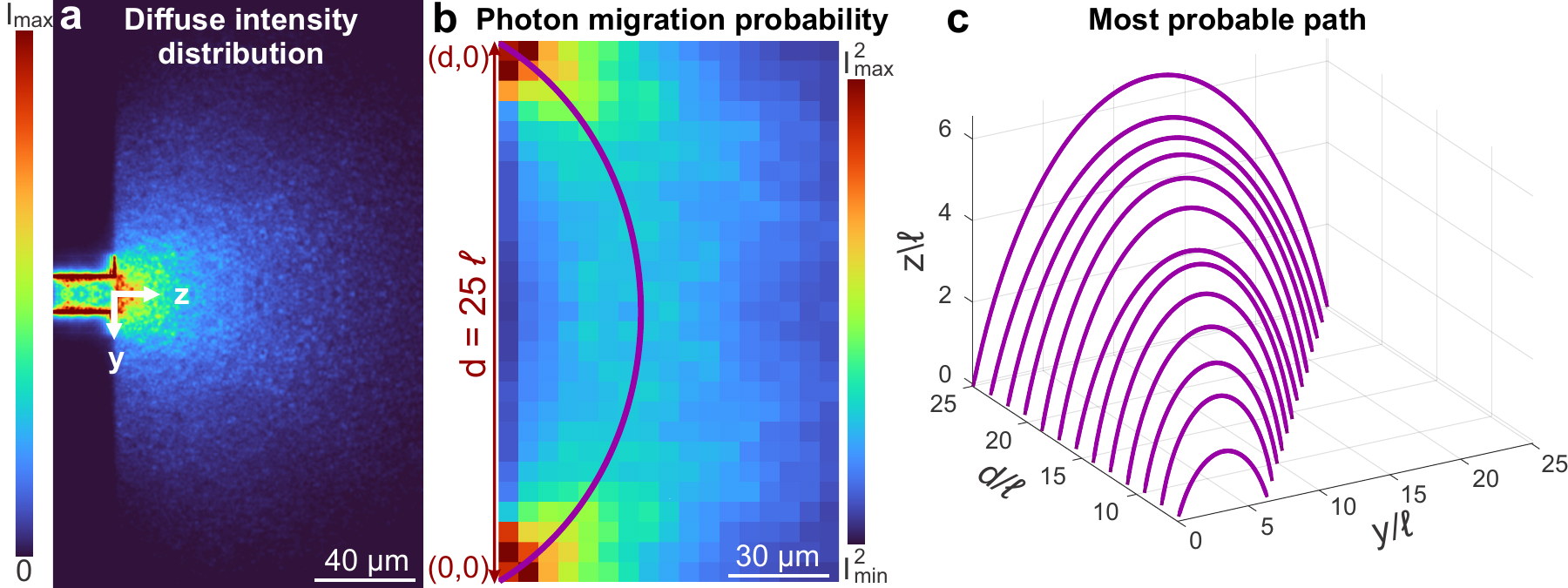}
    	\caption{\label{Sup:Figure3} 
    			{\bf Remission under random illumination.} (a) Ensemble-averaged intensity distribution inside the diffusive slab produced by random input wavefronts $\langle I_{\text{rand}}(y,z)\rangle$. 
    			(b) An example probability distribution of photon migration from the injection site at $(0,0)$ to the remission site $(d=25\ell, 0)$ via the position $(y,z)$ inside the diffusive slab. The probability is spatially averaged over $\ell$. The purple curve is the best fit of part of a circle to the location of maximal probability at each cross-section of $y$. It reveals the most probable path of the remitted light, referred to as the banana-shaped trajectory. 
    			(c) Most probable path of the remitted light as a function of the injection-remission distance $d$. In (a-b), data taken at 12 different wavelengths, in increments of 1 nm, from 1547 nm to 1558 nm are averaged.
    	} 
    \end{center}
    \end{figure}

    With the interferometric setup shown in Fig.~\ref{Sup:Figure2}, we can determine the field distribution across the disordered region of the sample for any incident wavefront. To measure the field of the light scattered out-of-plane from the sample, for a specific phase modulation pattern displayed on the SLM, we first measure the 2D intensity pattern of the out-of-plane scattered light across the entire disordered region by blocking the reference beam with a shutter. Then after unblocking the reference beam, we retrieve the phase profile of the out-of-plane scattered light with a four-phase measurement. In this measurement, the global phase of the pattern displayed on the SLM is modulated four times in increments of $\pi/2$ rad. By performing this measurement for a complete, orthogonal set of SLM phase patterns, we reconstruct a linear matrix $\mathcal{R}_{\rm SLM\rightarrow SLAB}$ that maps the field profile on the SLM to the field distribution across the entire slab. The remission matrix $\mathcal{R}_{\mathrm{SLM} \rightarrow d}$ is a part of $\mathcal{R}_{\rm SLM\rightarrow SLAB}$ that maps the fields from the SLM to the remission region. 
    
    Using $\mathcal{R}_{\rm SLM\rightarrow SLAB}$, we can recover the spatial field distribution anywhere inside the diffusive slab for any phase modulation pattern on the SLM. To check the accuracy of this recovery, we measure the spatial intensity distribution inside the slab and compare to that given by $\mathcal{R}_{\rm SLM\rightarrow SLAB}$. The measured spatial intensity distributions of a high-remission eigenchannel generated with phase-only modulation on the SLM have a Pearson correlation coefficient of $\sim 0.94$ with the corresponding intensity distribution predicted by $\mathcal{R}_{\rm SLM\rightarrow SLAB}$. 
    
    To experimentally generate different statistical ensembles, we average measurements of the same disorder configuration over different wavelengths of input light from a tunable laser. Specifically, 12 different wavelengths, in increments of 1 nm, from 1547 nm to 1558 nm are averaged to generate our experimental ensembles.

    \section{Theoretical model and prediction}
    
    \subsection{Filtered random matrix model}

    As explained in the main text, the $M_2 \times M_1$ remission matrix $\mathcal{R}$ is modeled as a filtered random matrix drawn from a larger  $M_0 \times M_0$ matrix $\mathcal{R}_0$ that characterizes transmission through an arbitrary disordered system supporting $M_0$ input and output propagating channels. This model captures the idea that only a fraction of the scattering channels are effectively excited and detected at the injection and remission sites. Both $M_0$ and the mean $\bar{\rho}_0$ of the bimodal eigenvalue PDF of $\mathcal{R}_0^\dagger \mathcal{R}_0$ are effective parameters of the model, which are unknown \textit{a priori}. A theoretical calculation of the flux and its fluctuations detected in the output port will be used to express $M_0$ and $\bar{\rho}_0$ in terms of the parameters involved in the remission experiment: the mean free path $\ell$, the injection-remission distance $d$, and the numbers of input and output spatial channels $M_1$ and $M_2$. 

The mathematical relation between the PDFs of $\mathcal{R}^\dagger \mathcal{R}$ and $\mathcal{R}_0^\dagger \mathcal{R}_0$ has been established in Ref.~\cite{2013_Goetschy}.  In our remission experiment $M_1>M_2$, so that $\mathcal{R}^\dagger \mathcal{R}$ has $M_2$ non-zero eigenvalues. The PDF of these non-zero eigenvalues, $ P(\rho)$, is equal to the eigenvalue PDF of $ \mathcal{R}\mathcal{R}^\dagger$. According to Ref.~\cite{2013_Goetschy}, it is given by $P(\rho)= -  \lim_{\eta \to 0^+} \mathrm{Im} [g(\rho + \textrm{i} \eta)]/\pi$, where $g(w)$ is the solution of the equation
\begin{equation}
\label{EqFRM}
\frac{w\,m_2g(w)+1-m_2}{m_2g(w)[w\,m_2g(w)+m_1-m_2]}g_0\left(\frac{\left[w\,m_2g(w)+1-m_2\right]^2}{m_2g(w)[w\,m_2g(w)+m_1-m_2]} \right)=1,
\end{equation}
where $m_1=M_1/M_0$, $m_2=M_2/M_0$ and 
\begin{equation}
\label{EqG0}
g_{0}(w)=\frac{1}{w}-\frac{\bar{\rho}_0}{z\sqrt{1-w}}\textrm{Arctanh}\left[
\frac{\textrm{Tanh}(1/\bar{\rho}_0)}{\sqrt{1-w}}\right].
\end{equation}
Equation~\eqref{EqFRM} expresses a relation between the Stieltjes transforms $g(w)$ and $g_0(w)$, from which it is possible to express the  moments of $P(\rho)$ in terms of those of the eigenvalue PDF of $\mathcal{R}_0^\dagger \mathcal{R}_0$. For the two first cumulants of $P(\rho)$, we find~\cite{2013_Goetschy}
\begin{align}
\left< \rho \right>&=m_1 \left< \rho_0 \right> = m_1\bar{\rho}_0,
\\
\text{Var}(\rho/\left< \rho \right>)&=m_2\left[\text{Var}(\rho_0/\left< \rho_0 \right>)  +\frac{1}{m_1}-1\right]=m_2\left[\frac{2}{3\bar{\rho}_0}  +\frac{1}{m_1}-2\right].
\end{align}
We now solve these equations to express the unknown parameters $m_1$, $m_2$, and $\bar{\rho}_0$ of Eq.~\eqref{EqFRM} in terms of the two first moments $\left< \rho \right>$ and $\left< \rho^2 \right>$. We get
\begin{align}
m_1&= \sqrt{\frac{3\left< \rho \right>}{2} \left[\beta \, \text{Var}\left(\frac{\rho}{\left< \rho \right>}\right) -1\right]+\left(\frac{3\left< \rho \right>}{2}\right)^2}+\frac{3\left< \rho \right>}{2},
\label{EqM1}
\\
m_2&=\frac{m_1}{\beta},
\label{EqM2}
\\
\bar{\rho}_0&=\frac{\left< \rho \right>}{m_1},
\label{EqMean}
\end{align}
where $\beta=M_1/M_2$.  By solving Eq.~\eqref{EqFRM} with above expressions for $m_1$, $m_2$, and $\bar{\rho}_0$, we are able to predict the full distribution $P(\rho)$ from the first two moments $\left< \rho \right>$ and $\left< \rho^2 \right>$ only.
Figure 4(a) of the main text has been obtained using this procedure. In addition, the upper edge $\rho_{\text{max}}$ of the PDF $P(\rho)$ is given by~\cite{2013_Goetschy}
\be
\label{EqSolMax}
\rho_{\text{max}}=w^*\left[1+\frac{m_1-1}{w^*g_{0}(w^*)}\right]\left[1+\frac{m_2-1}{w^*g_{0}(w^*)}\right] ,
 \ee
where $w^*$ is the solution of
\be
\left.\frac{\textrm{d}g_{0}(w)}{\textrm{d}w}\right\vert_{w^*}=\frac{g_{0}(w^*)}{2w^*}
\frac{-(1-m_1)(1-m_2)+w^{*2}g_{0}(w^*)^2}{(1-m_1)(1-m_2)-(1-m_1/2-m_2/2)w^*g_{0}(w^*)}.
\ee
Theoretical predictions shown in Figure 4(b) of the main text  have been generated by evaluating Eq.~\eqref{EqSolMax} for different injection-remission distances $d$ and mean free paths $\ell$.

\subsection{Theoretical evaluation of $\left< \rho \right>$ and $\text{Var}(\rho)$} 

\begin{figure}
    \begin{center}
    	\includegraphics[width=\linewidth]{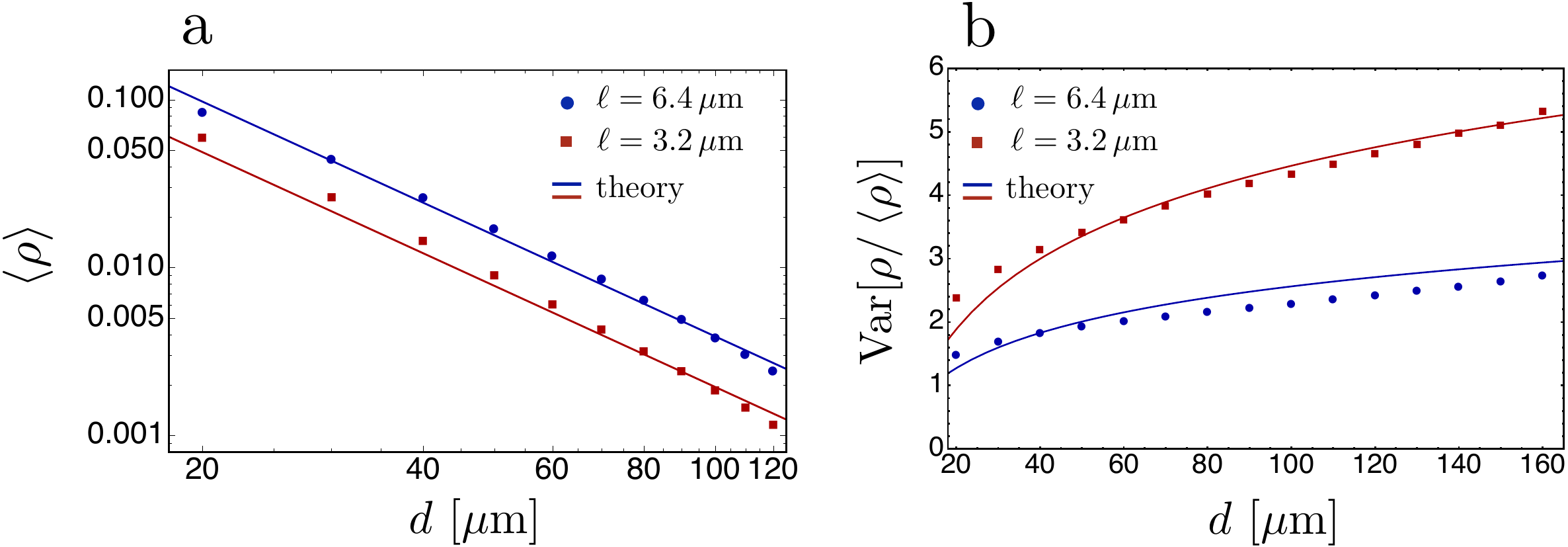}
    	\caption{\label{Sup:Fig_theory_MeanVar} 
    	{\bf Mean and variance of remission eigenvalues. }
    	Mean (a) and normalized variance (b) of the probability density function $P(\rho)$ of the non-zero remission eigenvalues. Symbols represent numerical data, and solid lines are predictions of our analytical model. With increasing injection-remission distance $d$, the mean $\langle \rho \rangle$ decreases as $1/d^2$, and the normalized variance Var[$\rho/\langle \rho \rangle$] increases as $\text{log}(d/W_1)$. 
    	} 
    \end{center}
    \end{figure}

Let us now evaluate analytically the first two moments $\left< \rho \right>$ and $\left< \rho^2 \right>$ of the PDF  $P(\rho)$. First, we note that $\left< \rho \right>= (M_1/M_2)\rho_{\text{rand}}$, where $\rho_{\text{rand}}$ is the mean diffusive  flux measured in the output waveguide, also equal to the mean $\left< \tilde{\rho}\right>$ of the eigenvalues $\tilde{\rho}$ of $\mathcal{R}^\dagger \mathcal{R}$. It is given by 
\be
\rho_{\text{rand}} \simeq W_2D\partial_z I(y= d, z=0),
\ee
where $I(y, z)$ is the solution of the stationary diffusion equation
 \be
 -D\nabla^2I(y, z)=\delta(y)\delta(z-z_\text{in}).
 \ee
 In previous equations, $W_2$ is the width of the output waveguide, $D=\ell c/2$ is the diffusion coefficient of light and $z_\text{in}$ is the depth at which the diffusion process is initiated (of the order of the mean free path $\ell$). The solution takes the form $I(y, z)\simeq z \ell/(\pi D y^2)$ in the limit $y \gg z,\ell $. Substituting $M_2=kW_2/\pi$, we get $\rho_{\text{rand}}\simeq M_2 k\ell/(kd)^2$ and
\be
\left< \rho \right> \simeq M_1 \frac{k\ell}{(kd)^2}.
\ee
This prediction is compared to numerical simulations in Fig.~\ref{Sup:Fig_theory_MeanVar}(a).

To find the expression of the normalized variance $\text{Var}(\rho/\left< \rho \right>)$, we first express it in terms of the full set of eigenvalues (including zeros) $\tilde{\rho}$ of  $\mathcal{R}^\dagger \mathcal{R}$ as
\be
\text{Var}\left(\frac{\rho}{\left< \rho \right>}\right) =\frac{M_2}{M_1}\left[\text{Var}\left(\frac{\tilde{\rho}}{\left< \tilde{\rho} \right>}\right) +1 \right]-1.
\label{EqVar1}
\ee
Note that $\left< \tilde{\rho} \right> = \rho_{\text{rand}}$. Then, we relate $\text{Var}(\tilde{\rho}/\left< \tilde{\rho} \right>)$ to the fluctuations of the total intensity $I_a$ measured in the output waveguide after exciting a channel $a$  of the input waveguide. This is done by means of the singular value decomposition $\mathcal{R}=U\hat{\rho}^{1/2}V^\dagger$, where $\hat{\rho}$ is the diagonal matrix $\text{diag}[\tilde{\rho}_n]$. Using this representation, we find $I_a=\sum_n \vert V_{an}\vert^2 \tilde{\rho}_n$. As a result, the fluctuations of  $I_a$ depend on the fluctuations of $\tilde{\rho}$ as well as the statistical properties of the matrix $V$. Assuming that $V$ is randomly distributed in the unitary group as in Ref.~\cite{2022_Bender_Deposition}, we get
\be
\text{Var}\left(\frac{\tilde{\rho}}{\left< \tilde{\rho} \right>}\right)\simeq M_1\text{Var} \left(\frac{I_a}{\left<I_a \right>}\right)
\label{EqVar2}
\ee
in the limit $M_1 \gg 1$. The average in the unitary group makes the result independent of the channel $a$. It is thus also equal to the fluctuations of intensity $I=\vert E \vert^2$ in the output waveguide resulting from a uniform excitation of all modes in the input waveguide. We evaluate the latter by decomposing the field $E$ as a sum of all possible scattering contributions reaching the output waveguide. Using standard diagrammatic techniques, we find
\be
\frac{ \text{Var}I}{\left<I \right>^2}=\frac{1}{M_2}+C_2,
\label{EqVar3}
\ee
with 
\be
C_2=\frac{1}{4k\ell\left<I(d,0)\right>^2}\iint dy dz \left<I(y,z) \right>^2\left[\nabla_{y,z} K(d,0;y,z) \right]^2,\label{eq:C2_2d}
\ee
where $\left<I(y,z) \right>$ is the mean intensity generated inside the disordered material at the position $(y,z)$ by a uniform illumination of the input waveguide located at the origin, and $K(d,0;y,z)$ is the Green's function of the stationary diffusion equation $-\nabla_{y', z'}^2K(y',z';y,z)=\delta(y-y')\delta(z-z')$, evaluated at the position $(d,0)$ of the output waveguide. The Green's function being equal to zero at the sample interface, it takes the form
\be
K(y',z';y,z)=\frac{1}{4\pi} \text{ln}\left[ \frac{(y-y')^2+(z-z')^2}{(y-y')^2+(z+z')^2} \right].
\ee
In the limit $d\gg W_1, \ell$, the calculation of the integral defining $C_2$ gives
\be
C_2 \simeq \frac{1}{k\ell} \left[\frac{4}{\pi} \text{ln}\left(\frac{d}{W_1} \right) +\gamma \right],
\label{EqC2}
\ee
with $\gamma \simeq (2\text{ln}2-7/2)/\pi +\pi/2 \simeq 0.9$. The value of $\gamma$ is sensitive to regularization procedure required to compute Eq.~\eqref{eq:C2_2d} and, therefore, may deviate in simulations from the value quoted above, while still remaining on the order of unity. Kwant numerical simulation of $C_2$ gives $\gamma\simeq0.6$. Finally, the combination of Eq.~\eqref{EqVar1}, Eq.~\eqref{EqVar2}, and Eq.~\eqref{EqVar3} yields
\be
\text{Var}\left[\frac{\rho}{\left< \rho \right>}\right] =\frac{M_2}{M_1}+M_2C_2. 
\label{EqVar4}
\ee
Numerical simulations of the eigenvalue variance shown in Fig.~\ref{Sup:Fig_theory_MeanVar}(b) confirm the theoretical result Eq.~\eqref{EqVar4}, which predicts a linear dependence on the mean free path as well as logarithmic dependence on the injection-remission distance $d$ . 

\subsection{Approximate FRM solution}

\begin{figure}
    \begin{center}
    	\includegraphics[width=0.7\linewidth]{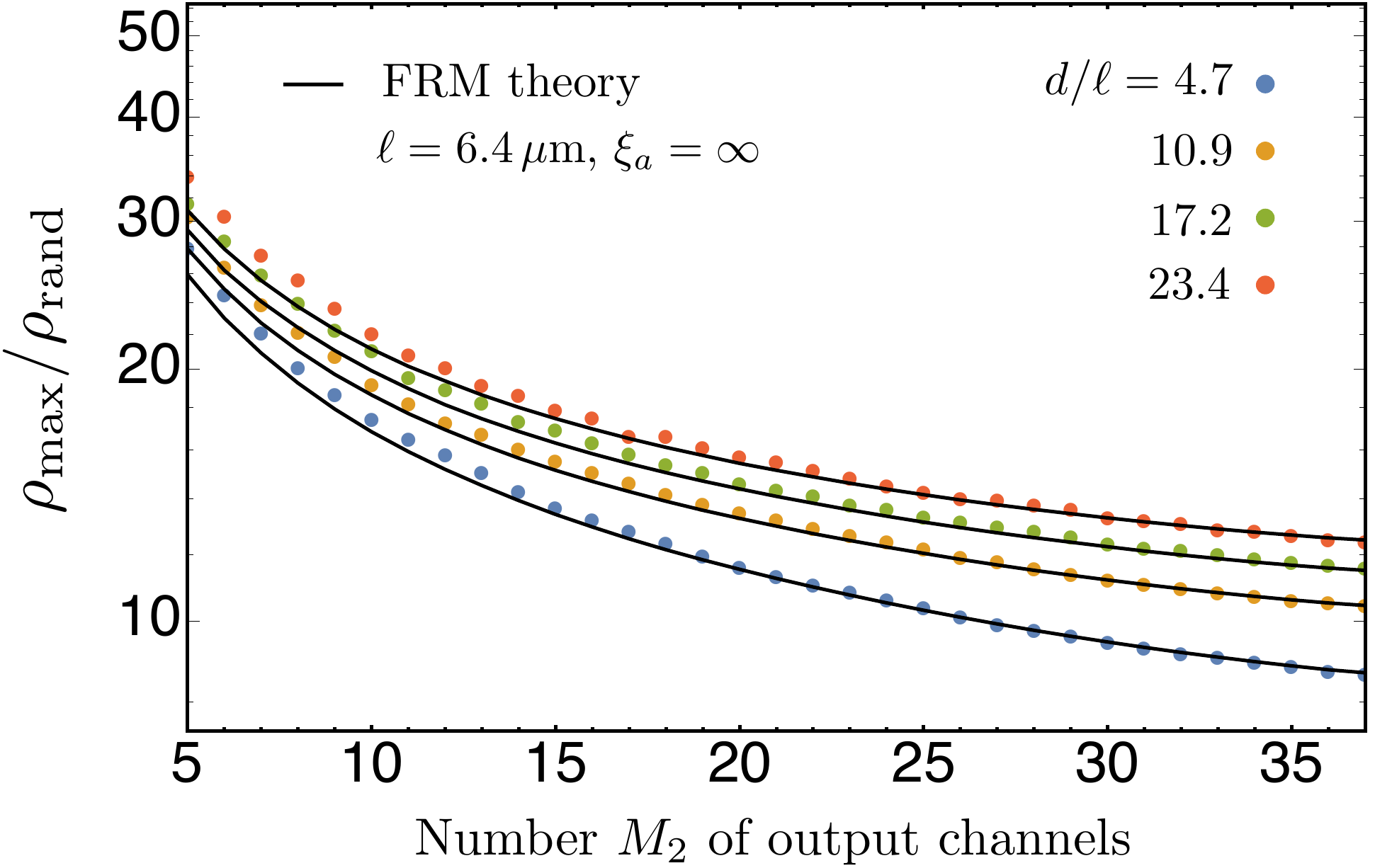}
    	\caption{\label{Sup:Fig_theory_M2} 
    	{\bf Dependence of remission enhancement on number of output channels. }
    	The maximal-remission enhancement $\rho_{\text{max}}/\rho_{\text{rand}}$ decreases with increasing the number of output spatial channels $M_2$. The number of input spatial channels is $M_1$ = 56. Analytical predictions (solid line) agree well with numerical results (symbols) for varying the input-output distance $d$. 
    	} 
    \end{center}
    \end{figure}

In order to gain some insight into the FRM predictions, we can perform an analytical expansion of the exact implicit solution given by Eq.~\eqref{EqFRM}. First, we note that for the distance $d$ covered in the experiment, the predicted PDF $P(\rho)$ is almost independent of the parameter $\beta = M_1/M_2$ appearing in Eq.~\eqref{EqM1} and Eq.~\eqref{EqM2}. Hence, we can take $m_1=m_2 \equiv m$ in these equations. We note that $P(\rho)$  still depends on $M_1$ and $M_2$ through $\left< \rho \right>$ and $\text{Var}(\rho/\left< \rho \right>)$. In addition, we also observe that $m<0.2$ for $d>40$ \textmu m,  which allows us to search for an approximate solution in the limit $m\to 0$. In this limit we establish in Ref.~\cite{2022_Bender_Deposition} that the upper edge of $P(\rho)$ takes the form
\begin{equation}
\label{SolEdge}
\frac{\rho_\text{max} }{\left<\rho\right>}\simeq \frac{[(\alpha-1)^{2/3}+(\pi/2)^{2/3}]^2[\alpha-1+(\pi/2)^{2/3}(\alpha-1)^{1/3}]}{\alpha(\alpha-1)^{1/3}}
+ \mathcal{O}(m),
\end{equation}
where $\alpha=m/\bar{\rho}_0$. Using Eq.~\eqref{EqM1} and Eq.~\eqref{EqMean} with $\beta =1$, we get
\begin{align}
\alpha& = \frac{m^2}{\left< \rho \right>}
\\
\nonumber
&= \left( \sqrt{\frac{3}{2} \left[\text{Var}\left(\frac{\rho}{\left< \rho \right>}\right) -1\right]+\frac{9\left< \rho \right>}{4}}+\frac{3\sqrt{\left< \rho \right>}}{2}
\right)^2
\\
\nonumber
&\simeq \frac{3}{2} \left[\text{Var}\left(\frac{\rho}{\left< \rho \right>}\right) -1\right]
\nonumber
\\
&= \frac{3}{2} \left(M_2C_2+\frac{M_2}{M_1}-1\right).
\end{align}
For the values of the mean free path  and injection-remission distance $d$ in our simulation and experiment ($\ell =6.4$ \textmu m or $\ell =3.2$ \textmu m, and $d>30$ \textmu m), the parameter $\alpha$ takes values  $\sim 3-6$. This indicates that the qualitative behavior of  $\rho_\text{max}$ can be captured by considering the limit $\alpha \gg 1$ of Eq.~\eqref{SolEdge}:
\be
\frac{\rho_\text{max} }{\left<\rho\right>}\simeq \alpha + 3\left(\frac \pi 2\right)^{2/3}\alpha^{1/3}-2 + \mathcal{O}(\alpha^{-1/3}).
\ee
In particular, in the limit $M_2C_2 \gg1$, this result predicts that the relative enhancement with respect to the random illumination should scale as
\begin{align}
\frac{\rho_\text{max} }{\rho_\text{rand}}&\simeq \frac{3}{2}M_1C_2
\nonumber
\\
&\simeq \frac{3W_1}{2\pi \ell} \left[\frac{4}{\pi} \text{ln}\left(\frac{d}{W_1} \right) +\gamma \right].
\label{EqMaxLeading}
\end{align}
This prediction captures the strong impact of the number of input channels and the mean free path, as well as the weak logarithmic dependence on the distance $d$, observed both in simulations and with the exact FRM solution (see Fig. 4b of the main text). We also note that the leading order given by Eq.~\eqref{EqMaxLeading} is independent of $M_2$ in the limit of large number of remitted channels ($M_2 \gg 1/C_2$). As $M_2$ is progressively increased, the enhancement factor first decreases and then saturates to a value larger than unity. This is confirmed in Fig.~\ref{Sup:Fig_theory_M2}, where we show the energy enhancement obtained by wavefront shaping as a function of $M_2$ for different injection-remission distance $d$. The completed dependence on $M_2$ is well captured by the solution of Eq.~\eqref{EqFRM}, while the saturation in the limit $M_2\gg1/C_2$ is qualitatively described by Eq.~\eqref{EqMaxLeading}.

In 3D, $C_2$ has been computed for transmission in the slab geometry in Ref. \cite{1989_Pnini}. For slabs of thickness $L\gg W_1$, it is found
\begin{equation}
    C_2^{(3D)}\simeq \frac{3}{ k^2 \ell W_1}\sim \frac{1}{k\ell}\times\frac{1}{\sqrt{M_1}},
    \label{eq:C2_3D}
\end{equation}
where $W_1$ is the diameter of the incident cylindrical beam. We note absence of any dependence on $L$ in transmission geometry. This result suggests that in 3D, even in remission geometry, $C_2$ correlations should saturate when distance from the source beam is large enough $d\gg W_1$. Therefore, we expect the transition between Gaussian and non-Gaussian limits for remission enhancement discussed in the main text (i.e. $M_2\sim 1/C_2^{(3D)}$), to take place for $M_2\sim(k\ell)\times M_1^{1/2}$ in 3D diffusive samples.
    
\subsection{ FRM predictions for samples with loss}

\begin{figure}
    \begin{center}
    	\includegraphics[width=1\linewidth]{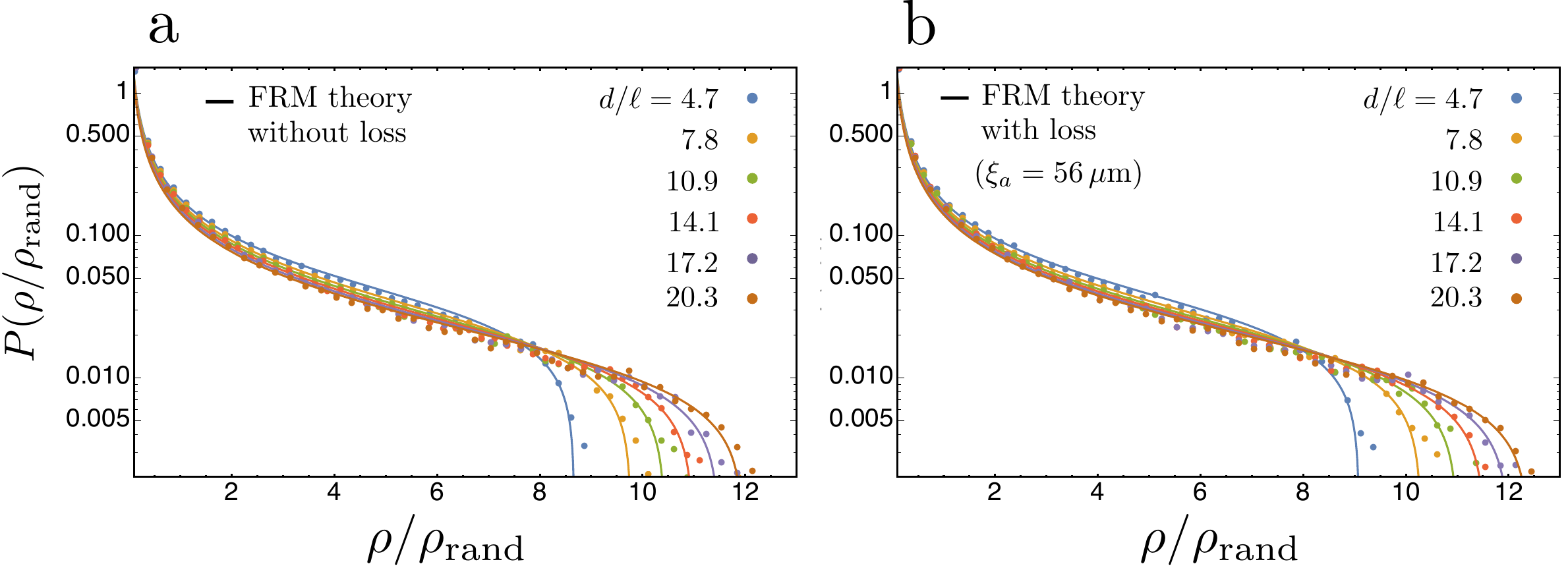}
    	\caption{\label{Sup:Fig_theory_Distribution} 
    	{\bf Remission eigenvalue distribution.}
    	Probability density of normalized remission eigenvalue $\rho/\rho_{\text{rand}}$ for varying the injection-remission distance $d$ normalized by the transport mean free path $\ell=6.4$ \textmu m in a 2D diffusive slab. In (a), the system has no loss, whereas in (b), the diffusive dissipation length is $\xi_a = 56$ \textmu m. Solid lines are theoretical predictions found by solving Eq.~\eqref{EqFRM}. 
    	} 
    \end{center}
    \end{figure}

\begin{figure}
    \begin{center}
    	\includegraphics[width=4in]{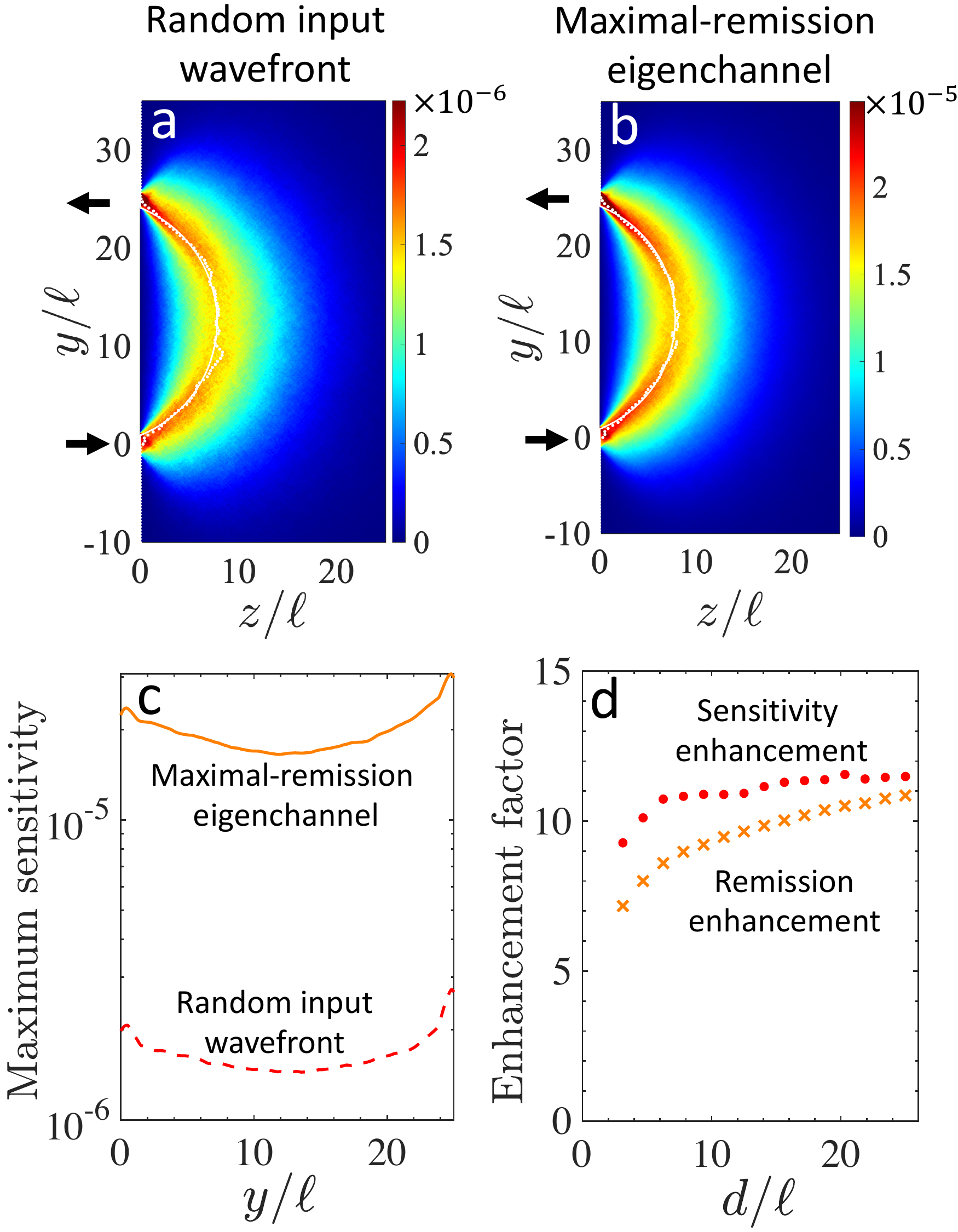}
    	\caption{\label{Sup:Figure4} 
    	{\bf Sensitivity enhancement by maximal-remission eigenchannel in a medium with loss.} 
        Numerically calculated sensitivity of remission due to local absorption inside a 2D diffusive system of uniform loss. The diffusive dissipation length $\xi_a$ = 56 \textmu m, all other parameters are identical to those of Fig.~\ref{Fig_sensitivity} in the main text. (a) for random input wavefronts, (b) for the maximal-remission eigenchannel. White dots denote the depth where the maximum sensitivity is reached for a given value of $y$, fitted by part of a circle displaced in the negative $z$ direction -- dashed line. Identical sensitivity maps in (a) and (b) confirm the penetration depth is not compromised by the enhanced remission.  
        (c) Maximum sensitivity vs. $y$ from (a,b) showing an order of magnitude enhancement by the maximal-remission eigenchannel. 
        (d) The sensitivity enhancement at $y=d/2$ (circles) compared to the remission enhancement (crosses) as a function of the injection-remission separation $d$.
    	} 
    \end{center}
    \end{figure}

The power and versatility of the FRM model can be further exemplified by analyzing the case of samples with loss. Let us consider the disordered system characterized by a diffusive dissipation length $\xi_a=56$ \textmu m, as in the experiment. The largest  injection-remission distance $d$ is three times the value of $\xi_a$, and the dissipation leads to a strong attenuation in the amount of signal $\rho_{\text{rand}}$ collected in the remission port under random illumination.  Here we evaluate the impact of loss on the full PDF $P(\rho)$. To set aside the effect of loss on the mean eigenvalue $\left< \rho \right> = (M_1/M_2)\rho_{\text{rand}}$, we normalize all non-zero eigenvalues $\rho$ of $\mathcal{R}^\dagger \mathcal{R}$ by $\rho_{\text{rand}}$, and present the resulting PDF in Fig.~\ref{Sup:Fig_theory_Distribution}. Remarkably, we find that losses have weak influence on the distribution $P(\rho/\rho_{\text{rand}})$, despite their strong impact on the mean $\left< \rho\right >$. A close comparison between Fig.~\ref{Sup:Fig_theory_Distribution}(a) and Fig.~\ref{Sup:Fig_theory_Distribution}(b)  reveals that the enhancement $\rho_\text{max}/\rho_{\text{rand}}$ is actually slightly larger in the presence of loss. Furthermore, we confirm that the FRM predictions given by Eq.~\eqref{EqFRM} and Eq.~\eqref{EqG0} still apply in the presence of loss, as solid lines faithfully capture the small changes found in the tail of $P(\rho/\rho_{\text{rand}})$ for arbitrary distance $d$.    
    
\section{Sensitivity Map}

Here we derive Eq.~\eqref{eq:sensitivity} in the main text. Consider scalar wave equation in 2D with ${\bf r} = (y,z)$. Let $E_t({\bf r}) = E_{\rm in}({\bf r}) + E_{\rm sca}({\bf r})$ be the unperturbed total field profile given $E_{\rm in}({\bf r})$ as the incident field with $ E_{\rm sca}$ satisfying an outgoing boundary condition. $E_t({\bf r})$ satisfies
\begin{equation}
\label{eq:wave_eq_unperturbed}
H_0 E_t({\bf r}) = 0, \quad
H_0 =  -\nabla^2 - k_0^2 \varepsilon_r({\bf r})
\end{equation}
where $k_0 = \omega/c$ and $\varepsilon_r({\bf r})$ is the relative permittivity profile of the disordered medium before perturbation.
Consider a small amount of absorption $d\varepsilon_i$ introduced to the imaginary part of the relative permittivity profile over a subwavelength area $A_{\varepsilon}$ centered at location ${\bf r}_0$, which we treat as a perturbation.
The incident field is fixed, and the total field changes to $E({\bf r}) = E_{\rm in}({\bf r}) + E'_{\rm sca}({\bf r})$, which satisfies
\begin{equation}
\label{eq:wave_eq_perturbed}
\left[ H_0 - i k_0^2 A_{\varepsilon} d\varepsilon_i \delta({\bf r}-{\bf r}_0) \right] E({\bf r}) = 0.
\end{equation}
Taking the difference between Eq.~\eqref{eq:wave_eq_perturbed} and Eq.~\eqref{eq:wave_eq_unperturbed}, we get
\begin{equation}
H_0 \left[ E({\bf r}) - E_t({\bf r}) \right] = i k_0^2 A_{\varepsilon} d\varepsilon_i \delta({\bf r}-{\bf r}_0) E({\bf r}_0).
\end{equation}
Defining the retarded Green's function $G({\bf r},{\bf r}')$ of the disordered medium by
\begin{equation}
H_0 G({\bf r},{\bf r}') = \delta({\bf r}-{\bf r}')
\end{equation}
with an outgoing boundary condition, we see that the resulting change in field profile is
\begin{equation}
\label{eq:dE}
dE({\bf r}) \equiv E({\bf r}) - E_t({\bf r})
= i k_0^2 A_{\varepsilon} d\varepsilon_i G({\bf r},{\bf r}_0) E({\bf r}_0)
= i k_0^2 A_{\varepsilon} d\varepsilon_i G({\bf r},{\bf r}_0) E_t({\bf r}_0),
\end{equation}
where we replace $E$ with $E_t$ on the right-hand side because the perturbation $d\varepsilon_i$ is small and we can invoke the Born approximation.

Next, we compute the resulting change in remission $R$. Define $R$ as the total power collected at the remission port divided by the incident power for incident field $E_{\rm in}$.
The time-averaged Poynting vector, $\langle {\bf S} \rangle = \langle {\bf E} \times {\bf H} \rangle$, is proportional to
${\rm Im} \left[ E^* \nabla E \right]$, so the remission is
\begin{equation}
\label{eq:R}
R = - \frac{\int_{\rm out} dy \, {\rm Im} \left[ E^*{\partial E}/{\partial z} \right]_{z=0}}{ \int_{\rm in} dy \, {\rm Im} \left[ E_{\rm in}^*{\partial E_{\rm in}}/{\partial z} \right]_{z=0}},
\end{equation}
where $\int_{\rm in} dy$ and $\int_{\rm out} dy$ indicate integration over the injection port and the remission port respectively.
Now, substituting Eq.~\eqref{eq:dE} into Eq.~\eqref{eq:R}, invoking reciprocity $G({\bf r},{\bf r}') = G({\bf r}',{\bf r})$, and keeping only terms to leading order of $d\varepsilon_i$, we obtain the change in remission as
\begin{equation}
\begin{aligned}
dR \equiv R - R_0 
&= - \frac{\int_{\rm out} dy \, {\rm Im} \left[ \frac{\partial (E-E_t)}{\partial z} E_t^* -(E-E_t)\frac{\partial E_t^*}{\partial z} \right]_{z=0}}{ \int_{\rm in} dy \, {\rm Im} \left[ E_{\rm in}^*{\partial E_{\rm in}}/{\partial z} \right]_{z=0}} \\
& = - k_0^2 A_{\varepsilon} d\varepsilon_i \frac{ {\rm Re} \left[ E_t({\bf r}_0) E_c({\bf r}_0) \right]}{ \int_{\rm in} dy \, {\rm Im} \left[ E_{\rm in}^*{\partial E_{\rm in}}/{\partial z} \right]_{z=0}},
\end{aligned}
\end{equation}
where
\begin{equation}
E_c({\bf r}) \equiv \int_{\rm out} dy' \left[ \frac{\partial G({\bf r},{\bf r}')}{\partial z'} E_t^*({\bf r}') -G({\bf r},{\bf r}')\frac{\partial E_t^*({\bf r}')}{\partial z'} \right]_{z'=0}.
\end{equation}
Invoking Green's theorem, we see that $E_c$ is the total field given $E_t^*$ in the remission port as the incident field\footnote{The evanescent components of $E_t$ in the remission port can either be included or excluded; it won't change the result since evanescent components do not contribute to the flux in Eq.~\eqref{eq:R}.}. 
This completes the proof of Eq.~\eqref{eq:sensitivity} in the main text.

    \newpage

      \newpage
 	
      \newpage
    \newpage

    \newpage


 \end{document}